\documentclass[10pt,a4paper]{article}
\usepackage{amsmath}
\usepackage{amsthm}
\usepackage{epsfig}
\usepackage{latexsym}
\usepackage{pdflscape}
\numberwithin{equation}{section}
\usepackage{graphicx}
\title{Religion-based Urbanization Process in Italy: Statistical Evidence from
Demographic and Economic Data \thanks{This paper is part of scientific activities in COST Action IS1104,
"The EU in the new complex geography of economic systems: models,
tools and policy evaluation". The authors are grateful to Dr. Eleonora Di Cristofaro and Dr. Antonio Caputo for helpful suggestions. All remaining errors are ours.}}
\author{ Marcel Ausloos$^{1,2,3}$
and Roy Cerqueti$^{4,}$\thanks{Corresponding author.}}
\date{$^1$ School of Management, University of Leicester\\
University Road, Leicester, LE1 7RH, UK \\
$^2$ e-Humanities, NKV, Amsterdam, The Netherlands \\ $^3$  Res. Beauvallon, rue de la Belle Jardini\`ere, 483/0021\\
B-4031, Li\`ege Angleur, Euroland \\ E-mail: marcel.ausloos@ulg.ac.be\\  $ $\\
 $^4$ Department of Economics and Law, \\University of Macerata, \\Via Crescimbeni, 20\\I - 62100  Macerata, Italy.\\ E-mail:  roy.cerqueti@unimc.it  }
\begin{document}
\maketitle
\begin{abstract}
This paper analyzes some economic and demographic features of Italians
living in cities containing a Saint name in their appellation
(hagiotoponyms). Demographic data come from the surveys done in 
the 15th (2011) Italian Census, while the economic wealth of such cities is explored through their recent
[2007-2011] aggregated tax income (ATI). This cultural problem is 
treated from various points of view. First, the exact
list of hagiotoponyms is obtained through linguistic and religiosity
criteria. Next, it is examined how
such cities are distributed in the Italian regions. 
Demographic and economic
perspectives are also offered at the Saint level, i.e. calculating
the cumulated values of the number of inhabitants and the ATI, "per
Saint", as well as the corresponding relative values taking into
account the Saint popularity. On one hand, frequency-size plots and
cumulative distribution function plots, and on the other hand,
scatter plots and rank-size plots between the various quantities are
shown and discussed in order to find the importance of correlations
between the variables. It is concluded that rank-rank correlations
point to a strong Saint effect, which explains what actually Saint-based toponyms
imply in terms of comparing economic and demographic data.

\end{abstract}
\textit{Keywords:} Econophysics, Italy, urbanization process, ranking, number of
inhabitants, tax income, correlation, hagiotoponyms.


\section{Introduction  }\label{sec:intro}
\par
The impact of religion on the demographic and economic evolution of
the societies has been clearly stated in several studies. In the
endless list of such contributions, one should mention the (not
recent but very interesting) survey on this topic of Iannaccone
(1998), but also Iannaccone (1991), Ellison and Sherkat (1995),
Lehrer (1995, 1999), Waters et al. (1995) and, more recently,
Caltabiano et al. (2006), Zhang (2008), Yeatman and Trinitapoli
(2008), Connor (2011) and Connor and Koenig (2015). The quoted
papers deal with the social dimension of the religion for what
concerns themes like marriage, education, population flows and
socio-cultural economics.
\newline
A field of research quite neglected by economists and
socio-scientists is that related to the influence of religion on the
urbanization process, along with an exploration of its historical
grounds economic and demographic characteristics.
\newline
It is indeed widely accepted that many cities have developed on
various seeds, e.g. water sources, river bridges, ores, ..., but
many exist due to some "religious purpose", in a large sense (see
Durkheim, 1968, Eliade, 1978, Dubuisson, 2003, Dennett, 2006, Stark
and Bainbridge, 1979). The cult allowed splitting, thus necessarily
juxtaposing,  the human experience of reality into sacred and
profane space and time.
\newline
In catholic religion, much religious activity  is performed through
the cult of Saints, considered as intercessors with  "God". Their
bones and relics could even make "miracles". The cult  of  Saints
emerged in the 3rd century and gained momentum from the 4th to the
6th century. It formed  from Greek and Roman veneration of
divinities, heroes, and rulers. It was established at "sanctified
locations" which in turn could  attract pilgrims, "priests",  and
merchants,  and thus did grow in population size, and became cities.
No need to say that it took many centuries before some strong
organization around the cult of some Saint developed. It is well
known that rival clergy up to bishops  and their cliques have been
fighting  lengthy battles over who had the right to claim the Saint
for their own community and even scorning "rival Saints". Whether
there were economic conditions in play  at those times  is not a
question raised here, -  the answer seeming obvious. One of the
causes  of the protestant reformed is  well known indeed.
\newline
In Italy (IT, hereafter),  known as a basically catholic country, -
being closely related to the siege of the papacy,
there has been for a while an important motivation about the cult of Saints,  
to the point that several cities bear the name of a Saint (Webb,
1996).
\newline
This paper explores the main characteristics of the Italian society
when isolating citizens living in the ''cities with a Saint name'',
i.e. hagiotoponym cities (see Reading, 1996). Specifically, this
religion-based relevant cultural aspect of the Italian urbanization
process is described under a demographical and economical point of
view. In particular, we aim at exploring what Saint-based toponyms
imply in terms of the comparison between economic and demographic
data.
\newline
One may thus first wonder how many  of these cities appeared and
where they are located. Also, it seems to be important to explore
how important are they with respect to the population size.
\newline
These questions can be justified as being related to studies on
social science and urban planning. Some motivation also arises from
touristic, geographical, and cultural points of view.
\newline
A third concern, but related to the previous one, is of linguistics
origin, i.e. a search for the statistical distribution of names of
Saints attached to city names. The reader is addressed to the
Dictionary of Farmer and Hugh (1987). A related question concerns
the rare Saints, e.g. those occurring only once, so called hapaxes;
how many of them? The question "why are such Saints rare?" is
outside the purpose of the present report. Nevertheless, how they
geographically distributed is an original question.
\newline
A  fourth motivation of our investigation, but not the least, is
based on economics concern. How does being a hagiotoponym city,
behave  with respect to others? Moreover, is a specific Saint
"better", in an economic sense, than another?
\newline
Therefore, the first question being tackled in this report is: what
city and how many, in IT, have a ''Saint'' name in its appellation?
The (not so trivial as it could be thought, at first) methodology is
explained in Sect. \ref{sec:method}.  In particular, the data
acquisition needs some very careful work as explained in Sect.
\ref{sec:data}. Note that a distinction will be made between male
and female Saints. In Sect. \ref{sec:Data treatment}, we define the
economic and demographic quantities which have retained our
attention.
 \newline
Moreover, it seems of complementary interest to observe the
occurrence of less frequent Saints, e.g. dislegomena Saints. Indeed,
one might have asked, at the time of rising of such cities whether
some novelty in names (or cult), some rarity or in {\it a contrario}
some popularity of a Saint, would have been (and is still)
beneficial  within some "competition".
\newline
In studies of cities, one often examines  the rank-size or
size-frequency relationships, in accord to Vining (1977) and Chen
(2012). Their analysis will be the  main content of the  figures
displayed here below in Appendix B and commented in Sect.
\ref{sec:analysis}. For what concerns the rank-size analysis, a
brief comment is needed. Often, Zipf's law (Zipf, 1949) is able to
fit the relationship between the size and the rank of a given
variable. Such a regularity has not a strong theoretical ground (see
Krugman, 1995), and it is shown not  to provide a robust analysis of
the connections between demographic and economic data in our
specific context. In this respect, it is important to note that
Zipf's law fails in several cases (see e.g. Giesen and Sudekun,
2011; Soo, 2007; Cordoba, 2008; Garmestani et al., 2008; Bosker et
al., 2008; Vitanov and Ausloos, 2015). For this major reason, we
have considered more general rank-size rules based on the
Lavalette's law (Lavalette, 1966). In doing this, we have found
robust best fits results, with high levels of $R^2$ and interesting
interpretations. More specifically, in order to  observe how many
Saints (or cities) concern the above questions, plots for the
rank-size, size-frequency and cumulative distribution function can
be made and are presented in order to find an appropriate empirical
law: see Sect. \ref{sec:analysis}.
\newline
The $density$ of Saints, with respect to the whole country,   and
with respect to regions  should give some idea on ''$catholicity$''
(or  some $cult$)  "touristic spreading", at the time of community
creations. Thus, whether there is any geographical distribution,
with special Saints in some area, is discussed as well; see Sect.
\ref{sec:analysisregionalfrequency}.
Next, since cities have often developed around churches or chapels,
the question is raised  whether there is any relation between
''cities with Saint names''  (hagiotoponyms) and present city
population size.  Moreover, are the most popular Saints associated
with ''large'' cities? Is the cumulative distribution of interest?
In so doing, one may next wonder about the wealth of such cities.
How do they fare nowadays? This is examined through the (modern)
aggregated income taxes (ATI) of such cities -- which represents the
aggregate contribution that the citizens of each city provide to the
national Gross Domestic Product (GDP) -- in Sect. \ref{sec:regATI}.
The final answer to these questions is found  through rank-rank
correlation studies,  in Sect. \ref{sec:rankrank}.
\newline
Section \ref{sec:analysis} contains also a comparison between the
overall IT cities situation and the set of hagiotoponyms in the
respect of demographic and economic variables. It seems that the
hagiotoponym cities build a reality similar to the IT one, when the
biggest cities are removed as outliers. We strongly emphasize the
condition or constraint.
\newline
Note that there is no discussion here below concerning parishes nor
churches nor chapels nor folk life events implying Saints. This
demands  a very complex survey; thus is not studied here.  In this
respect, an interesting paper is Kim (2007) on pilgrimage and towns
in medieval christianity.  We recommend its reading both for the
outlined ideas but also for the bibliography. Nevertheless, we
stress that pilgrimage  towns do not necessarily have a Saint for
name. Cities we examine here below  are in fact very rarely
pilgrimage  towns, though they have local Saint activities.
\newline
Moreover, questions on why religious cities grow or fail, and why
several Saints are more popular than others are left for further
work; see Stark (1996) for a starting point.
\newline
A brief conclusion is found in Sect. \ref{sec:conclusion}. Appendix
A contains the needed statistical toolkit employed for the analysis
of the data, while Appendix B contains the Figures and Tables
presented.

\vskip 12pt \textbf{Remark.} When finishing writing this paper, we
became aware\footnote{through $medievalist.net$} of a University of
Birmingham  2009 Master of Philosophy thesis, by C.H. Morris
(Morris, 2009) 
who examined quite related topics, i.e. the practice of venerating
holy figures and their relics, and events that surround such
worship, in Anglo-Saxon and Medieval England. Indeed,  more or less
quoting Morris' thesis abstract:
\begin{itemize}
\item[] {\it ''[The cult of Saints] was a cultural phenomenon that engaged all sections of society, and Saints
enjoyed high levels of popularity through their cults. Not all
instances were the same, and cults differed in size and
construction. The distances over which cults could command attention
varied, as did the social groupings that they catered for.''}
\end{itemize}
This very interesting work differs from ours, surely about the
respectively concerned lands, England $vs.$ Italy,  the timing, and
scientific approaches, but is highly complementary.  We emphasize
that we add much  to this sort of investigations by considering also
economic questions.

\section{Construction of the dataset}\label{sec:method}
In IT, there are 8092 cities or, better, \textit{municipalities} (in
Italian: \textit{comuni}) shared among 20 regions nowadays. Thus,
data  on population size and on ATI have been collected at the
municipality level, before selecting  the cities of interest.
\newline
Specifically, we  want to identify the municipalities which have a
toponym related to a Saint, so called $hagiotoponyms$.  Sometimes,
it is not so easy to understand who is the truly related Saint. For
example,  Giovanni  could be the ''apostle'' or the ''baptist''.
However, we consider that he is most likely Giovanni the apostle.
Indeed, the baptist is now called ''Giovanni Battista'' of
''Gianbattista''.  A check of the official list of Saints on Farmer
(1987), indicates that
Giovanni can be associated also to
more recent people, but this is irrelevant to a city
original appellation. 
The identification of the true Saint is sometimes not possible.
Therefore, we decided   neither  to pursue nor explore  such a task
further.  The Saint name in our list will point  to a unique Saint,
who  is then considered to be  the representative element of the
related category of the Saints with the same name. It might be
relevant to remove this constraint in more advanced religious
studies.  We also understand that a Saint having given his/her name
to a city is not necessarily a {\it bona fide} catholic Saint, but
only a "Saint by tradition" (see Farmer, 1987). We do  not consider
this  ambiguity relevant to our consideration, - on the contrary.

\subsection{Data acquisition}\label{sec:data}

\begin{figure}
  \includegraphics[width=5.8in]
  {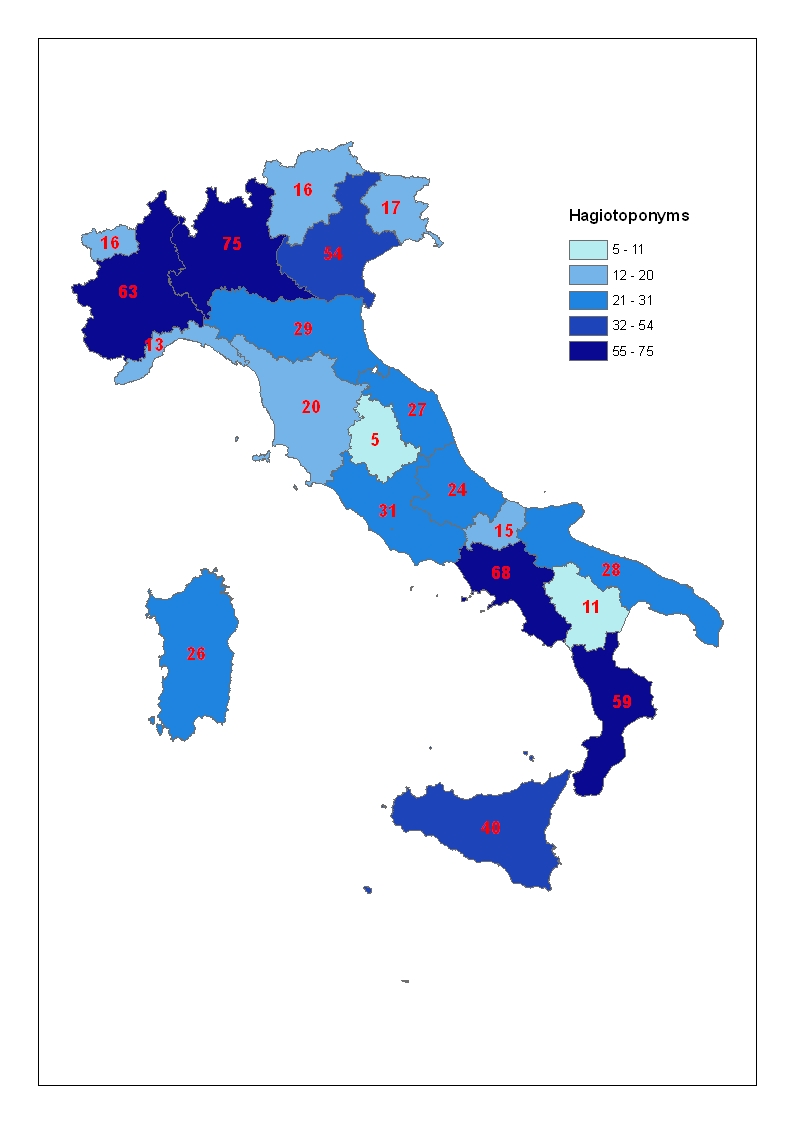}
\caption{Map of Italy with the regional distribution of the
hagiotoponyms.} \label{fig17:Italy}
\end{figure}

The investigated data is of three different natures: population data
and economic data, and the city names.
\begin{itemize}
\item
The source of the population data is the Italian Institute of
Statistics (ISTAT). In particular, data on the population are
extracted from the elaborations of the 15th Italian Census,
performed by ISTAT in 2011.
\newline
The population data taken for the municipalities are: number of
inhabitants, males and females, number of families, people living in
a family, average number of the components of a family, people
living as cohabitant and not as a family. Not all such
statistical data can be examined here with respect to our concerns,
but we recommend the data sources for subsequent work. 
\item
The economic data, i.e. aggregated  tax income (ATI), was obtained
from (and by) the Research Center of the Italian Ministry of
Economics and Finance (MEF). It covered the 2007-2011 time window.
The number of cities in IT, and their regional or provincial
membership have changed during this time interval, - in fact going
from 8101 to 8092. In order to quantify the cities from some
economic point of view, we have averaged the ATI of each city over
that time window (denoted as $<ATI>$), taking into account the
official merging of specific cities, when
necessary. 
\newline
A warning is in order: the number of hagiotoponym cities has
remained constant  in the considered  time interval.  However, it
was found that two hagiotoponym cities  changed regions: San Leo and
Sant'Agata Feltria moved from Marche to Emilia Romagna in 2009.
However,    the available official data considers such an
administrative change as appearing in 2008. To be scientifically
consistent, we agree with the  official dataset and consider such
municipalities as belonging to Marche in 2007 and Emilia Romagna
after 2007.
\item
The data collection here above mentioned  relies  on the identification of Italian cities
having a toponym recalling a Saint. Such an identification has been a tedious stage. It was
implemented in several  phases which seemed worthwhile of presentation for justifying the subsequent data. Firstly, a preliminary list of cities has been
constructed through the application of four sorting procedures.
Secondly, a refinement of the preliminary list has been applied, and
some municipalities have been ejected on the basis of our own established
criteria. After this second phase, the  list of municipalities
with ''Saint'' appellation has been revised one by one. We  make more precise here below the
applied procedure.
\newline
The preliminary list moves from the premise that some specific
toponym might come out from deformations of the original Saint name.
Therefore, the string ''san'' -which, at a first look, seems to be
rather informative, being actually contained also in the words
''santo'', ''santa'', ''sant'', - would in fact lead to the removal
of a number of acceptable municipalities from the full list of
cities (like \textit{Camposampiero}, in Veneto, which derives from
\textit{San Pietro}). Thus, we provide a first sorting procedure by
employing the string ''sa'' to the entire collection of 8092
elements. The resulting list contains 1321 municipalities. The
second sorting procedure has been attained by employing the strings
''San '', ''Santo '', ''Santa '', ''Sant' '' (note the blank space
at the  end of the strings) for the list of 1321 municipalities.
This second sorting has produced 636 municipalities containing at
least one of the strings and the remaining 685 ones without the
strings. The third sorting procedure has been implemented on the
group of 685 municipalities, in view of further facilitating the
identification of the municipalities of interest. The string ''san''
(without space after the word) has been employed; whence it is found
that 180 municipalities contain ''san'' and the other 505 without
the string. The two groups of 180 and 505 have been manually checked
line by line. All such cases have been carefully examined. In
particular, as expected, there exists a number of municipalities
whose toponym is a linguistic transformation of a Saint name. We
call them \textit{strange cases}. As an example, \textit{Samugheo},
in Sardegna, derives from \textit{San Michele}. In presence of a
strange case, the assessment of the (possible) corresponding Saint
name has been performed through the reading of historical
information related to the single controversial municipalities. This
information has been taken from the official website of the
municipalities and/or from Wikipedia. When no information is
available, the candidate strange case has been removed from the
preliminary list of municipalities (like \textit{Santadi}, in
Sardegna). All accepted  as a valid hagiotoponym but "strange cases"
are listed in  Table \ref{tab:strange-cases}.

Hence, in the groups with and without the string ''san'', a number
of 28 and 23 municipalities, respectively, have been selected for being contained in
the preliminary list. Some \textit{collateral effects}
came out from checking the municipalities after the third sorting
procedure. First of all,   \textit{Madonna del Sasso} has been
included in the preliminary list, \textit{Madonna} being  an Italian
name for the "Virgin Mary", - a human Saint. We became also aware
about the existence of some municipalities in Valle d'Aosta with
French Saint names and in Trentino Alto Adige with German Saint
names. Thereafter,  we performed a fourth sorting procedure in the total
list of 8092 municipalities, by employing the strings ''donna'',
''dame'', ''Frau'' to search for  the equivalent of the Virgin Mary not only
in Italian, but  also in French and German, respectively. The fourth
sorting procedure gave us the possibility to add to the preliminary
final list 1 further municipality not already included,
\textit{Rhemes-Notre-Dame}.  The preliminary list, so, collects
636+28+23+1=688 municipalities.

Thereafter, came the treatment of several ambiguities, leading to
 several municipalities  exclusions. The criteria for exclusion
have been basically  due to  (i) names which do not  "obviously"
point to a specific human Saint and/or (ii)  whether the toponym,
though being a "sanctified location",  is clearly derived from  some
Bible fact or event. Thus, we have not considered as "Saints"  51
municipalities containing "san" as follows:
\begin{enumerate}
\item Acquasanta Terme
\item Abbasanta
\item Villasanta
\item Villa Santina
\item Luogosanto
\item Camposanto
\item Lagosanto
\item Pietrasanta
\item Sant'Arcangelo - santarcangelo (3 times)
\item Sant'Angelo - santangelo (24 times)
\item Santa Croce (5 times)
\item  Santa Luce (Terme)
\item Sansepolcro
\item Santopadre  
\item (Borghetto) Santo Spirito
\item San Salvatore - San Salvo (6 times)
\item San Buono

\end{enumerate}
Some further explanation (or argument for rejection) about  a few of
the elements in the  above list can be given.
\newline
Items 1.-8. point to something which has been sanctified (some
examples: 1. \textit{acqua}-\textit{water}; 5.:
\textit{luogo}-\textit{place}; 6.: \textit{campo}-\textit{field};
7.: \textit{lago}-\textit{lake}). Items 9. and 10. refer to the
general "concepts" of angel and archangel. Item 11. derives from the
Holy Cross of the martyrization of Christ. Item 12. could be
confused with \textit{Santa Lucia} but should not be. Item 13. is a
linguistic transformation of  Santo Sepolcro: it seems that
this city was originating from a chapel built by Egidio e Arcano, in
memory of the Jerusalem Holy Sepulchre, and does not refer to a
human Saint. Items 14.-16. refer to the Holy Trinity. \ Santo
Padre  is the Italian way to denominate the Pope, but here it refers
directly to the christian God. Santo Spirito stands for the
Holy Spirit of the Trinity, while Salvatore (or, less
frequently, Salvo) is the usual Italian appellation of
Jesus, - who is not a Saint. Item 17. denotes a village which was
named Sancti Boni  or  Castrum Bonum  at the time of
its foundation, to revere the holy goodness. However, it does not
point to a specific Saint, and has been removed from the preliminary
list.
\newline
Nevertheless, we kept \textit{Michele}  (11 times) and \textit{Raffaele}  (once) as $bona$ $fide$ Saints,
although they are not humans, but archangels. However, they are so much
anthropomorphic that they can be here assimilated to human Saints.
\newline
A few remarks are in order.  After the constitution of the final
list, we have assigned the related Saint to each of the 637
municipalities. The sum of the frequencies of Saint appellations is,
unexpectedly, 639. Indeed, there are two municipalities whose
toponym contains a couple of Saints (\textit{Santi Cosma and
Damiano} in Lazio and \textit{San Marzano di San Giuseppe} in
Puglia). For these particular cases, the available municipality
data, $both$ in terms of population size and economic features, has
been shared equally between the two Saints of the couple.
\newline
Linguistic transformations have been treated case by case. As an
example, \textit{San Lorenzello} can be identified with \textit{San
Lorenzo}. Therefore, they belong to the same class
(\textit{Lorenzo}'s one). In several ''difficult'' situations, the identification procedure has been analogous to that of the strange cases, i.e. the reading of the historical notes about the municipalities. The case of  Monte San Pietrangeli, in Marche, is a nice example. We did not find any Saint named
 Pietrangeli. The historical notes provide a different name of such a municipality, still used by its inhabitants, which is \textit{Monsampietro}. The reference Saint is then \textit{Pietro}, the first Pope.
\newline
Such linguistic transformation cases have been collected in Table \ref{tab:ling-trans}.
\newline As hinted here above, other complex cases are the municipalities
with a hagiotoponym written in a foreign language. There are some
(16) French names in Valle d'Aosta and some (9) German names in
Trentino Alto Adige. For the German names, it is a case of
application of the bilingualism of that Region, and there is a
(legal) Italian counterpart (translation). The adopted criterion has
been to translate, when possible, the French names in Italian, and
take only the available translation in Italian of the German names.
Results are shown in Table \ref{tab:strangers}. \noindent The number
of different  Saints, in the final list of municipalities, is  thus
206, as reported in Table \ref{tab:Tabla2} and Table
\ref{tab:Tabla5},  containing 31 females and 175 males\footnote{We
have decided that archangels are males.}; in which 15 and 101 are
attributed to only 1 city (hapaxes); see Table \ref{tab:Tabla5}.
\end{itemize}
The resulting distribution of the hagiotoponyms at a regional level
can be found in Figure \ref{fig17:Italy}.

\subsection{Data treatment}\label{sec:Data treatment}
To exemplify the procedure, consider that we made two\footnote{The
Tables were either 637 or 639 long items, depending whether we
considered real cities or hagiotoponyms} Tables: one with the number
of inhabitants and one with the $<ATI>$ values, both in decreasing
order, according to the Saint name, distinguishing between males and
females.  The top and bottom of each Table is shown in Table
\ref{tab:Tabla3}, \ref{tab:Tabla4}, \ref{tab:Tabla3-ATI}, and
\ref{tab:Tabla4-ATI} in view of outlining the number ranges. Table
\ref{TableATIstat} contains the statistical indicators computed for
$<ATI>$ and population data.
\newline
To highlight the role played by each Saint, population and economic
raw data have been further treated in two ways.
\begin{itemize}
\item Data have been cumulated over the name of the Saints as
follows:
$$
ATI(x)=\sum_{j \in F(x)}ATI_j, \qquad POP(x)=\sum_{j \in F(x)}POP_j,
\qquad \forall\,x,
$$
where $x$ denotes the name of a Saint. Thus, $F(x)$ is the set of
municipalities which  toponym derives from the Saint $x$, and
represents the popularity of $x$. Specifically, the cardinality of
$F(x)$ -denoted as $|F(x)|$- is the $frequency$ of $x$. $ATI$ and
$POP$ are the intuitive notations for the ATI values and for the
number of inhabitants datum, respectively. In particular, $ATI_j$
and $ATI(x)$ ($POP_j$ and $POP(x)$) represent the ATI (number of
inhabitants) datum for the municipality $j$ and the Saint $x$,
respectively.

\item
Furthermore, all data have been  cumulated and next averaged  to
take into account  the name ($x$)  popularity $F(x)$  of the Saints:
$$
\overline{ATI}(x)=\frac{1}{|F(x)|}\cdot \sum_{j \in F(x)}ATI_j,
\qquad \overline{POP}(x)=\frac{1}{|F(x)|}\cdot \sum_{j \in
F(x)}POP_j, \qquad \forall\,x.
$$
\end{itemize}

\section{Statistical analysis of the data} \label{sec:analysis}
First of all, we provide a description of the dataset through the
main statistical indicators. Then, some specific aspects of the
considered dataset will be treated, on the basis of the
methodological techniques described in the Appendix.

\subsection{Main statistical indicators}
The computation of the main statistical indicators leads to some
interesting outcomes:
\begin{itemize}

\item it appears that there are 91  cities with a $female$ Saint
name  for 31  (female) different names,  as reported in Table
\ref{tab:Tabla2} and Table \ref{tab:Tabla5}; the most popular is
Santa Maria  (including Marie)  occurring 23 times, much preceding
Sant'Agata (12 times).
\newline
Our statistical analysis shows that the distribution is rather
skewed  (skewness $ \sim 3.63$) and the kurtosis  $\sim 13.39$;


\item from Table 5 and Table 6, it appears that there are 546
cities with one (or two)  $male$ Saint name(s). The most popular is
San Pietro  (43  times), but San Giovanni (36 times) is not far off.
There are 175 different names. (Recall restrictions, due to Cosmo
and Damiani, and to Marzano and Giuseppe, as if there were 548
different cities).
\newline
For these 548 hagiotoponyms, our statistical analysis shows that the
distribution is rather skewed  (skewness $ \sim 4.57$) and the
kurtosis   $\sim 23.93$;

\item there are 206
different Saints  (175  males   +  31 females)  within the above
grouping rules for 639 hagitotoponyms.  This popularity ($F$)
distribution is rather skewed  (skewness  $\sim 4.55$) and the
kurtosis $\sim 24.0$.
\end{itemize}
The values of the statistical indicators provided in the above list
illustrate an hagiotoponym-type of urbanization mostly related to a
few very popular Saints (Pietro and Giovanni for males, Maria and
Agata for females), but with a very large number of cities recalling
unfrequent Saints. This is totally in line with the deep cult of
very important and popular religious figures like the Virgin
(Maria), the first Pope (Pietro), etc., along with widespread local
traditions related to less famous Saints.

Table \ref{TableATIstat} collects the main statistical indicators
related to the raw ATI and number of inhabitants data. This Table
can be compared with the main statistical indicators associated to
ATI and number of inhabitants when the entire set of IT cities is
considered (Table \ref{Tablestat-IT}). It is evident that the
minimum values of ATI and number of inhabitants share the same
magnitude order in the overall IT and hagiotoponym cases, while the
maximum ones are remarkably different (the maximum of IT is much
greater than that of hagiotoponyms for both the considered
variables). However, IT cities are on average slightly richer and
more populated than hagiotoponyms. Further information is bought by
other statistical indicators: IT cities are noticeably more volatile
than the set of hagiotoponyms (higher variance either for population
and ATI), and exhibit also a greater level of kurtosis.
\newline
All these facts point to hagiotoponyms which describe a hypothetical
IT situation when the main outliers are removed, both for the number
of inhabitants as well as for the ATI.

Table \ref{tab:stat-treated-data} collects the main statistical
indicators for the quantities $ATI(x)$, $POP(x)$,
$\overline{ATI}(x)$, and $\overline{POP}(x)  $ with respect to the
independent variable $x$.
\newline By a methodological point of view, the closest integer has been taken for $\overline{POP}(x)$.
Moreover, at these levels, and thereafter, we have not distinguished
between male and female names.  The ratios between the number of
cities having a hagiotoponym  or the relative number of different
Saints (in both case $\sim 0.17$)  or the proportion of  female
Saints in the overall counting (in both cases $\sim 0.14$) are not
small. Nevertheless, it is not expected that distinguishing genders
would bring much to the discussion. Moreover,  it should be obvious
to the reader that to take into account the gender would lead to
triple the number of curves, figures, columns in Tables, or Tables.
However, we do not disregard the interest of such an investigation
in the future,  for  a complementary paper.

\subsection{Saints frequency regional disparities} \label{sec:analysisregionalfrequency}

\begin{figure}
  \includegraphics  [height=14.8cm,width=14.8cm]  {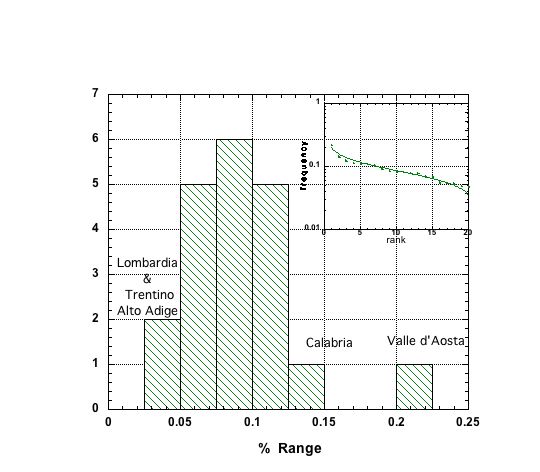}
\caption{Histogram of the relative number of hagiotoponym  cities
with respect to the total number of cities in the 20 Italian
regions, stressing the peculiarity of Valle d'Aosta, as an outlier
and the more extreme cases Lombardia, Trentino-Alto Adige and
Calabria. In an insert, the rank-size relationship is shown for the
20 regions, with a fit by the Lavalette function, Eq.
(\ref{Lavalette2}); the parameter values are: $\kappa_2$ = 0.082;
$\chi$ = 0.285; $R^2$ = 0.955.}\label{Plot7histogfig1ins}
\end{figure}

It can be observed  from  Table 2 that, on average, there are $\sim
8\%$ hagiotoponyms in the Italian regions. However, Valle d'Aosta is
an outlier in the hagiotoponym distribution, since about 22\% of the
municipalities contain a Saint name in that region. Recall that
Valle d'Aosta is an autonomous-like region, with much historical
connections with France.  In fact, in France, more than 5000
municipalities, out of  about 35000, i.e. about $15\%$, contain a
Saint name.
\newline
In  the more extreme Italian cases Lombardia and Trentino-Alto Adige
are "very poor" in hagiotoponyms:   $\sim 5\%$, though these are
regions in the North of IT, like Valle  d'Aosta.
\newline
In contrast,  Calabria, a southern region,  has a much above average
$\sim 14\%$ hagiotoponym content.
\newline
Note that the largest percentages of female hagiotoponyms occur in
Umbria (0.6666), Sicilia (0.375)  and Calabria (0.1864),  as for the
female/all ratio.
\newline
In contrast, the   largest percentages of male/all hagiotoponyms
occur in Friuli-Venezia Giulia (0.9412) and in  Lazio and Valle
d'Aosta (0.9375).
\newline
 Basilicata has a noticeable feature: there is no female Saint  name in any city.
\newline
{\it In fine}, from the search of an empirical law point of view,
the finite size effect of the data is remarkably emphasized when
searching for the hagiotoponym frequency distribution; see insert of
Fig. \ref{Plot7histogfig1ins}. Taking such a finite size into
account, the rank-size relationship for the 20 regions can be well
reproduced by a fit  with a  function as Eq. (\ref{Lavalette2}). To
our knowledge, the fundamental reason for the validity of such a
function, in this type of considerations, is unknown,  but rather
{\it ad hoc}. Only mathematically plausible arguments (Naumis and
Cocho, 2008) are known, but they seem hardly applicable in our case.

\subsection{Saints frequency empirical distributions} \label{sec:analysisempiricaldistribution}

\begin{figure}
  \includegraphics[width=5.8in] {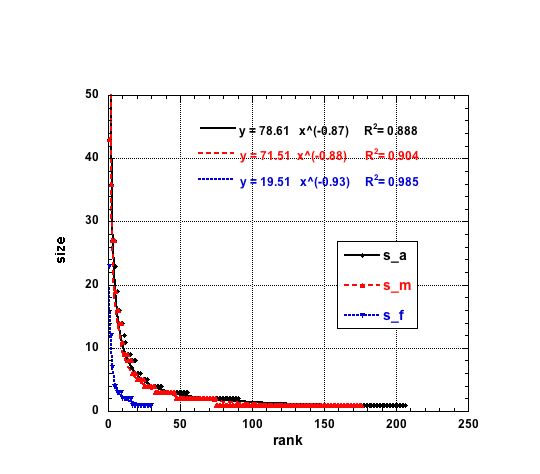}
\caption{Display of the rank-size relationship for  cities bearing a
Saint name ranked in decreasing order according to their  size, with
corresponding power law fits as indicated; the gender (f or m) is
distinguished beside the overall (a)  size (s) data. The best merely
hyperbolic fits are indicated.} \label {Plot21fig2lili3pwlf}
\end{figure}

It has been shown here above that there are 206 different Saint
names. Thereafter, a rank-size (Zipf) plot on classical axes, Fig.
\ref{Plot21fig2lili3pwlf},   can  be presented, i.e., the number of
cities, independently  carrying the name of a specific Saint, ranked
in decreasing order of the Saint popularity.  In so doing, we are
only focussing on a linguistic-like approach, as a function of the
rank, i.e. how many times the Saint occurs (its "size") in
hagiotoponym cities. The display, for the various genders and the
whole data set, indicates a smoothly decreasing data, as if    a
Zipf law exists.
\newline
However,  according to  Fig. \ref{Plot21fig3loloZM}, displaying the
same data as on  Fig. \ref{Plot21fig2lili3pwlf},  but on a log-log
plot, the rank($r$)-size($s$) relationship for  the number of times
a city has the name of a Saint ($i$ $\sim$ male (m), female  (f) or
all (a) cases) is obviously seen to be hardly represented by a mere
power law. A more appropriate fit is through a Zipf-Mandelbrot law,
Eq. (\ref{ZMlikeCr}), with downward curving at low rank. The
parameter values are given in the Figure. Nevertheless, note the
slight \textit{king effect} for Maria -i.e.: the distortion effect
due to the outlier Maria, see   Laherr\`ere   and Sornette (1998)-
($\nu \le0$). Note also that the power law decay at high rank is
very similar for each gender, with an exponent ($\zeta$) close to 1.
Such fits are rather remarkable since the regression coefficient
$R^2 \ge 0. 99$.
\newline
A  log-log display of the frequency-size relationship, Eq.
(\ref{V2}), i.e. the frequency of the size, for  the Italian cities
bearing a Saint name,   is shown in Fig.
\ref{fig2:Plot2frqsizelolo}. The corresponding fits with a
Zipf-Mandelbrot law, Eq. (\ref{ZMlikeCr}),  are indicated; the
gender ($f$ or $m$)  is distinguished beside the overall    (a) size
(s)  frequency data. The fits are rather remarkable, since $R^2 \ge
0.  99$.  It should not be surprising in such a plot to observe a
slight king effect for the male case, nor the largest value of the
$\zeta$ exponent in the all Saint case, due to the small influence
of the (rare) high size Saints.
\newline
A final plot, in examining the given Saint size distributions is
through a  log-log display of the  cumulative distribution  function
(CDF) as a function of the "size" relationship,  Eq. (\ref{V3}),
i.e. Fig. \ref{fig3:Plot41CDFlolo}.   Recall that   cities bearing a
Saint name are listed  in decreasing order of  their frequency. The
regression coefficient is again very high for  fits with  a
Zipf-Mandelbrot law, Eq. (\ref{ZMlikeCr}). A technical point is in
order concerning the  female data fit. The latter is very unstable
due to the small number of points, i.e. 7. The parameter values much
depend on the initial conditions imposed in the Levenberg-Marquardt
algorithm\footnote{ For completeness, note that the
Levenberg-Marquardt algorithm (see Levenberg, 1944, Marquardt, 1963,
Lourakis (2011) has been used  for the fitting procedure of the data
to the mentioned non-linear functions. The error characteristics
from the fit regressions, i.e. $\chi^2$, d, the number of degrees of
freedom, the $p-value$,  beside the $R^2$ regression coefficient,
have been calculated, but are not shown for space saving.   It  has
been observed that in all cases the $p-value$ is lower than
$10^{-6}$. }. This is due to the large number of approximately
equivalent minima in the parameter space; this unavoidable fact is
well known (see Herzel et al., 1994, Goldstein et al., 2004, Clauset
et al., 2009, Rawlings et al., 1998).

\subsection{Population size considerations}\label{sec:Stcitysize}
The number of inhabitants in the 639 hagiotoponym cities   is
presented on a  linear-linear plot in Fig.
\ref{fig9:Plot1NinhabitStrlili639}. Visually, this  looks like
displaying a smoothly, hyperbolic-like, decaying data, with an
exponent close to 1. However, the  $R^2$ value  is pretty low, i.e.
$\sim 0.54$.  Alas,  as  better seen in Fig.
\ref{fig8:Plot1NinhabitStrlolo3folo},  there is no nice  simple fit,
by an empirical law with few free parameters.
\newline
Indeed,  Fig. \ref{fig8:Plot1NinhabitStrlolo3folo} presents a
log-log  plot for the rank-number of inhabitants in the 639 Italian
hagiotoponym cities. Visually, from the data scattering, it cannot
be expected that a simple empirical law can be found: four  simple
laws are indicated, but do not lead to a convincingly interesting
regression coefficient. The Zipf-Mandelbrot law has a $R^2\sim0.98$,
but is far from being visually appealing at high rank. It  can be
concluded, at this stage, that the sampling is far from a random
one.
\newline
Note that these simple fits  for the  "all Saints" case (Fig.
\ref{fig9:Plot1NinhabitStrlili639} and Fig.
\ref{fig8:Plot1NinhabitStrlolo3folo}) are not nice enough to suggest
a decomposition between males and females  in further work.
\newline
In view of the change in curvature of the data near the middle of
the rank range, it is inappropriate to consider a fit by a power law
or any other purely convex function. Instead, like for Fig. 1 insert
case, it seems that a  fit by a 3-parameter function with inflection
point, as that given in Eq. (\ref{Lavalette3a}) in Appendix, is more
appealing. This is shown in Fig. \ref{figA1:Plot3Nihab639lilo4Lav3},
on a semi-log display of the number of inhabitants in the 639
Italian cities wearing a Saint name,  - cities ranked in decreasing
order of the number of inhabitants;   a fit by such a  function
shows a convincing $R^2\sim0.992$. Moreover, the fit for $r\le350$
is quite visually appealing.
\newline
These results can be compared with what the overall 8092 IT cities
say.
\newline
There is evident presence of outliers at a high rank, as the
histogram in Fig. \ref{Plot 1histoNinhab} noticeably puts in
evidence. This fact is further confirmed by the fits for IT, which
seem to be more of high quality when outliers are removed. In this
respect, the comparison between the best 3-parameters Lavalette fit
in the two cases of all 8092 cities and of removal of the 80 highest
rank cities is pretty informative, being the latter more visually
appealing than the former (see Figures \ref{Plot 4Lav3-80lolo} and
\ref{Plot4Lav3-80lilo}).
\newline
Also the case of low rank is quite interesting. Fig.
\ref{Plot1Lav3lowrlilo} shows that the low rank IT cities are nicely
fitted by a 3-parameters Lavalette curve, with $R^2 \sim 0.986$.
\newline
To conclude, by ranking cities with respect to the number of
inhabitants, the sample of hagiotoponym cities behaves closely to
the overall IT cities when outliers are removed.

\subsection{Economic considerations} \label{sec:regATI}

The $<ATI>$ of all the 639 Italian cities containing a Saint name
over the period 2007-2011 has been displayed on log-log axes in Fig.
\ref{fig8b:Plot6ATI5lolo639fits4}. It is seen on this figure that
simple laws, as those tested, i.e. power, exponential, log,  do not point to a plausibly simple
empirical relationship.
\newline
In fact, in view of the change in curvature of the data near the
middle of the rank range, it is inappropriate to consider a fit by a
power law or any other purely convex function. Instead, like for
Fig. 1 insert case, it seems that a  fit by a 3-parameter function
with inflection point, as that given in Eq. (\ref{Lavalette3a}), is
more appealing. A semi-log display of the $<ATI>$ the 639 Italian
hagiotoponym cities, cities ranked in decreasing order of their ATI
is shown in Fig. \ref{figA2:Plot6ATI5lilo639fitsLav1}.   A fit by
such a 3-parameter free  function shows a convincing fit for
$r\le350$, and an acceptable  $R^2\sim0.989$. The latter is smaller
than in the case of the population size, in Fig.
\ref{figA1:Plot3Nihab639lilo4Lav3}, but the exponents seem quite
similar.







Also in this case, the comparison between the outcomes of the
economic analysis of the hagiotoponyms and the one related to the IT
cities might be of some usefulness.
\newline
At this level, we refer the reader to Cerqueti and Ausloos (2015),
where rank-size rules are applied at a national as well as at a
regional level for cities in IT. The size is given by the ATI, and
data are disaggregated on the basis of municipal unit.
\newline
Cerqueti and Ausloos show that IT national economic data seems to be
well described by a 3-parameters Lavalette curve, even if the
distortion effect due to the presence of outliers is of high
magnitude. This is the so-called \textit{king and vice-roy effect},
see Section 4.2 in the cited paper. For what concerns the validity
of Zipf-Mandelbrot law, the hagiotoponym sample behaves according to
the overall IT and to the majority of IT regions, in that it
statistically fails in all the cases, being Lazio region a
remarkable exception.
\newline
Hence, we can reasonably state that the set of hagiotoponym cities
proxies IT cities without outliers, when ATI is considered in the
context of rank-size rules.

\subsection{Study of the correlations }\label{sec:rankrank}
In view of the above findings, considering that similar empirical
laws seem to hold for relations between the city population, Fig.
\ref{figA1:Plot3Nihab639lilo4Lav3}, and the city wealth, Fig.
\ref{figA2:Plot6ATI5lilo639fitsLav1}, as a function of the rank in
the relevant variable, it seems of interest to  search  whether such
ranks are correlated.  Such an answer is  obtained from so called
scatter plots (see Bradley, 2007), which allows to have some insight
in the correspondence between data lists when the measures
themselves are of less interest than their relative ordering
importance.
\newline
First, a log-log display of the scatter plot for the $<ATI>$ and the
number of inhabitants, for   the 639  Italian hagiotoponym cities,
is presented in Fig. \ref{fig11a:Plot1scatterNinhATIlolopw}. The
best power law fit indicates a loosely compact set of points along a
quasi linear function. A correlation seems plausible, but there are
a few outliers.
\newline
Next, the corresponding scatter plot for the cumulated  variables,
i.e. $ATI(x)$ and $POP(x)$, thus for the 206   Saints, can be
observed in Fig. \ref{fig11b:Plot2scattNinhagATIaglolopw} on log-log
axes. Again, a fine power law with an exponent close to 1 is found.
\newline
Third, Fig. \ref{fig11c:Plot3scattNinhxATIxlolopw} is the
corresponding log-log display of  the  scatter plot of the averaged
over 5 years  ATI and the number of inhabitants, in cities
corresponding to   the 206   Saints,  but  when reduced  as a
function of the frequency (popularity) of the Saint, i.e.
$\overline{ATI}(x)$ and $\overline{POP}(x)$. Again the  exponent of
such a power law fit is close to 1.
\newline
The  regression coefficient is the highest  for the cumulated data,
as should be expected.  However, the $R^2$ is much lower ($\sim0.8$)
in the latter case, indicating much scattering, whence a rather
strong  deviation from a  perfectly correlated popularity
(frequency) effect.
\newline
In order to quantify some correlation in  two  (necessarily equal
size) sets,  e.g., between population and ATI data, the Kendall's
$\tau$ rank measure (Kendall, 1938) is usefully calculated.
\newline
First of all, it is important to note that $N$= 206 implies $p+q  =
21115$, when there is no overlap, being $p$ and $q$ the number of
concordant and discordant pairs, respectively. However, in the
present case, two hapax Saints (Bassano and Sosti) have the same
cumulated number of inhabitants (2209), whence $p+q$= 21114.
\newline
The relevant data, i.e. the Kendall $\tau$, Eq. (\ref{taueq}),  and
correlation statistics of ranking order   between (i)
$\overline{POP}(x)$ and    $\overline{ATI}(x)$;  (ii) $POP(x)$  and
$F$; (iii)  $ATI(x)$  and $F$; (iv) $\overline{POP}(x)$  and $F$;
(v)       $\overline{ATI}(x)$  and $F$, for 206  Saints  ($x$), with
notations as in the text,  are given in Table
\ref{TabletaurankATIPOPF2ndtype}. For a visual inspection of the
correlations among the variables belonging to this set of data, of
some usefulness can be Figs.
\ref{fig11e:Plot2scattNinhagATIaglolopw} and
\ref{fig11f:Plot3scattNinhxATIxlolopw}.
\newline
The marked variations between the various $\tau$ coefficients allow
interesting observations. First of all, one can compute the Kendall
coefficient between the number of inhabitants and $<ATI>$ for the
637 hagiotoponym cities, and obtain $\simeq 0.81$. Admitting that
there might likely be some different wealth regime of the
inhabitants in the various cities, it is  {\it bona fide} expected
that the total $<ATI>$ of a city would be somewhat in direct
(simple) relation with the number of inhabitants. In fact, there are
variations in the $<ATI>/N$ (ATI per inhabitants) values. For this,
we found: mean $\sim 9813$; median $\sim 9593$; standard deviation
$\sim 3893$. However the relative pair ranking concordance ratio
$p/q$ is $\sim 9.7$. It maybe concluded that there is a high
concordance.
\newline
It is here interesting to point out the analogies and disparities
among hagiotoponyms and IT when dealing with the Kendall rank
correlation, being the variables under scrutiny ATI and number of
inhabitants.
\newline
The Kendall $\tau$ for the case of IT cities is reported in Table
\ref{Tabletaurank-IT}. It is immediate to check that Kendall $\tau$
is substantially the same in the case of hagiotoponyms ($\tau \sim
0.849$) and IT ($\tau \sim 0.850$). From this, it can be concluded
that there is a strong regularity in the correlations between the
two types of city ranking in IT. Moreover, it is also important to
stress that the discrepancies between IT and hagiotoponyms can be
found in the number of inhabitants and in the ATI, and not in how
the corresponding ranks are associated. This further confirms what
already found for the relationship between IT and Saint cities.
\newline
The matter is quite different  for the  cumulated data rank
correlations observed at the Saint level:  $\tau \simeq 0.850$ or
$\tau \simeq0.788$, for which $p/q\sim  12.35$  or $p/q\simeq 8.44$,
either for the total cumulated data per Saint of for the reduced
value taking into account the Saint popularity: columns (i) and (ii)
in Table \ref{TabletaurankATIPOPF2ndtype}. This huge variation in
$p/q$ is, surprisingly, pointing to a redistribution of the ranks,
following some sort of randomization.
\newline
The effect is amplified when the correlations with the Saint
popularity are  examined: in this case, $\tau \sim 0.510$ for
$POP(x)$, with  the   ratio $p/q \sim 3.08$, and a similar $\tau
\sim 0.518$ is found for the rank correlation between  $ATI(x)$ with
respect to the frequency $F$ of a Saint. In the latter case, the
ratio $p/q \sim 3.15$ (see columns (iii) and (iv) in Table
\ref{TabletaurankATIPOPF2ndtype}). Therefore, it can be concluded
that the cumulated population of cities for a given Saint is far
from being positively correlated with the Saint popularity.
\newline
When taking into account the Saint popularity and the cumulated data
of cities into the Saint level, we have that the ranking
correlations with respect to the Saint popularity leads to a very
small $\tau\sim 0.068$ or $\tau \sim 1.02$. Note that the ratios
$p/q$ are $\sim 1.15$ and $\sim 1.22$; see columns (v) and (vi) in
Table \ref{TabletaurankATIPOPF2ndtype}. From a purely statistical
perspective, this result is very close to prove an independence of
the rank sets.
\newline
Finally,   on one hand, this indicates how careful one should be in
drawing conclusions from one statistical indicator only.  On the
other hand, it points to a   "Saint" effect, either positive or
negative, depending on the considered variable.


%



 \section{Conclusion} \label{sec:conclusion}

In summary, let us note that the objectives of the study  were  to
describe the Italian society when referring to a key aspect of it:
the cult of catholic Saints and its reflection on the toponyms of
the Italian cities.  With this aim, we have found that:
\begin{itemize}
\item[(0)]
there exists an exhaustive list of cities in IT whose toponym is a
derivation of a human Saint name. This requested some linguistic
approach beside some religiosity filter. The resulting dataset can
be used for subsequent studies dealing with related problems;

\item[(i)]
there exists a rather simple empirical law for  the distribution of
Saint Names as hagiotoponym of cities, in particular in IT, and who
are the hapax Saints;

\item[(ii)] there exists a rather
simple empirical law about the distribution of population sizes for such cities;

\item[(iii)] there exists a rather simple empirical law about the
wealth distribution,  for such cities, measured through their
Aggregated Income Tax;

\item[(iv)] there exists some correlation between such data.
Specifically, there is a high concordance between the ranking of the
average ATI and of the population for the hagiotoponyms.

\item[(v)] there are qualitatively reasonable causes
which can be given for the findings.

\end{itemize}

The empirical laws are not trivial ones, thereby  proposing further
mathematical  investigations on them.  This remark  holds for the
city population and the  ATI. Correlations exist between both
variables, but with some loose ties because of the outliers.
\newline
Our conclusion on city population $vs.$ ATI correlations tends to
indicate that these cities with hagiotoponyms are not drastically
different from those in the rest of IT. In particular, it seems that
hagiotoponym cities may represent a proxy of the overall IT cities
when outliers are removed, both at an economic as well as a
demographic level. Moreover, the rank-rank relationship between ATI
and number of inhabitants -- obtained by employing the Kendall
$\tau$ -- leads to a robust outcome of concordance. The results on
Saint popularity, {\it in fine}, seem to indicate a lack of
correlations between the latter and the two main variables which we
have examined.
\newline
On (v), further explanations are needed. Let us distinguish females
and male names. The most important cult to have developed in
christianity is that of the Virgin Mary, whence St. Maria is
naturally an appealing Saint name for providing some cult in some
city, and having given the name to several places.
\newline
Yet, it seems interesting that she is less popular than Pietro and
Giovanni, who both were the closest apostles to Jesus, according to
christian tradition. Interestingly Martino comes third, before
Giorgio. Martin is also very popular in France and other countries,
where his name is usually more popular (Ausloos, unpublished) than
Giovanni  (Jean) and Pietro (Pierre). Giorgio's 4th place is
interesting: his name occurs much in IT, but also all over Europe.
The popularity of George is likely due to the religious role played
by Saint George, who killed dragons which synthesized the devil and
 its hell.
\newline
Agatha, as the second most popular female Saint is  also
interesting. She represents one of the most important martyr of the
sicilian christianity: she is venerated at least as far back as the
sixth century, "because" she had her breasts cut off, whence of
interest for cults by women in order to produce gynecological
miracles.  Why men, usually leading the populations at those early
Christian times, would give her name to a city is nevertheless an
open question.

\section*{Appendix A: Methodological instruments}
This Appendix contains a few lines on the  theoretical background of
the empirical laws, along with an explanation of the Kendall $\tau$,
for the reader convenience.
\subsection*{Formulas for fit}
\begin{figure}
  \includegraphics[width=5.8in]  {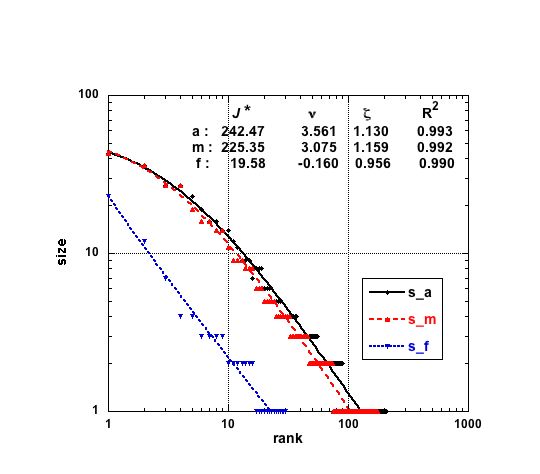} 
\caption{Log-log display of the rank-size relationship for  cities
bearing a Saint name ranked in decreasing order according to their
"size",  with corresponding  Zipf-Mandelbrot law fits as indicated;
parameters are for Eq. (\ref{ZMlikeCr}); the gender (f or m)  is
distinguished beside the overall   (a)  size (s) data. Note the
slight king effec for Maria ($\nu \leq 0$) in the "female data".}
\label{Plot21fig3loloZM}
\end{figure}

\begin{figure}
  \includegraphics[width=5.8in]  {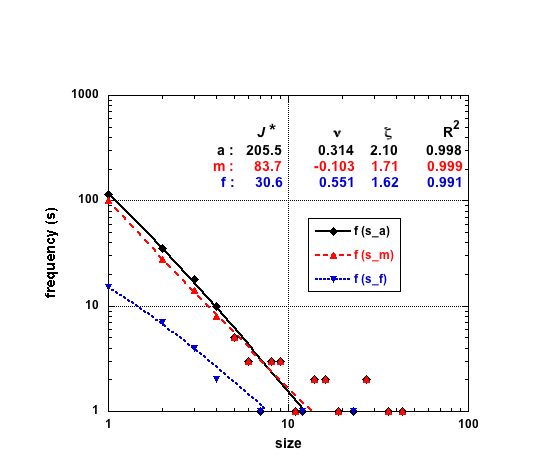} 
\caption{   Log-log display of the frequency-size relationship for
cities bearing a Saint name,  with corresponding Zipf-Mandelbrot law
fits as indicated; parameters are for Eq. (\ref{ZMlikeCr});  the
gender (f or m)  is distinguished beside the overall    (a)  size
(s)  frequency data. The most popular Saint name, in the context, is
Pietro, see Table 5. Observe the convex curvature for the  male data
cumulative distribution function, with a slight king effect for
males, due to the large number (101) of hapax Saint males, i.e. for
the "small size" region.} \label{fig2:Plot2frqsizelolo}
\end{figure}

\begin{figure}
  \includegraphics[width=5.8in] {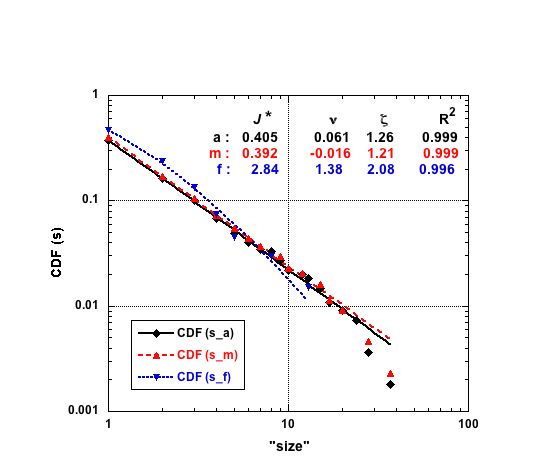}
\caption{    Log-log display of the  cumulative distribution
function (CDF) as a function of their "size" for  cities bearing a
Saint name in decreasing order of  their  "frequency", with
corresponding Zipf-Mandelbrot law fits as indicated; the gender (f
or m) is distinguished beside the overall    (a)  size (s) data.}
\label{fig3:Plot41CDFlolo}
\end{figure}

\begin{figure}
  \includegraphics[width=5.8in]  {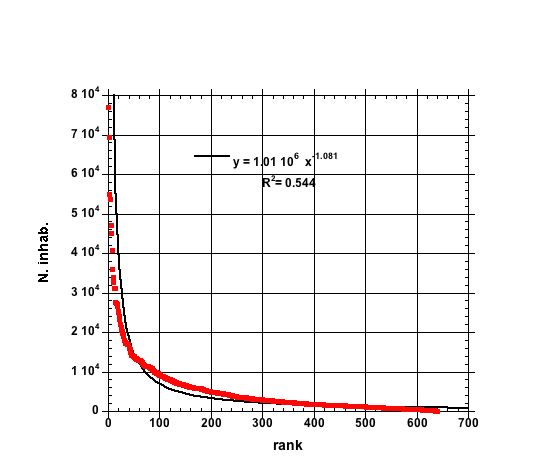}
\caption{Classical axes  display of the rank-size (number of
inhabitants in the  639 (637 + 2; see text)
 Saints generating Italian hagiotoponyms)
ranked in decreasing order of the  number of inhabitants. A simple
power law fit is indicated. Visually, this plot  looks like
displaying a smoothly decaying data, but  the regression coefficient
$R^2$ is pretty low; alas,  as "better" seen in Fig.
\ref{fig8:Plot1NinhabitStrlolo3folo}, it is not possible to find a
"nicely simple" fit.} \label{fig9:Plot1NinhabitStrlili639}
\end{figure}

\begin{figure}
  \includegraphics[width=5.8in]
  {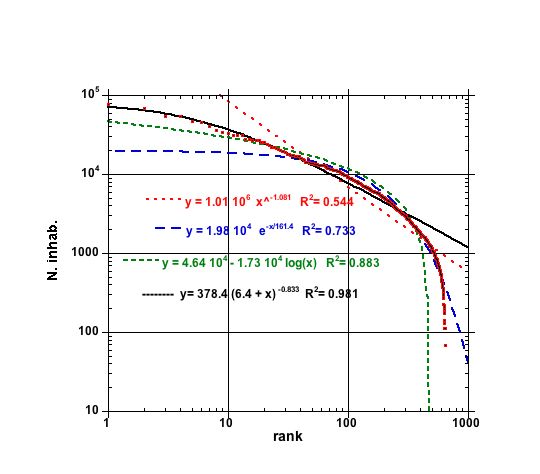} 
\caption{Log-log display of the Number of inhabitants in the 639
(637 + 2; see text) Saints generating Italian hagiotoponyms,  -
cities ranked in decreasing order of the number of inhabitants; four
simple law fits are shown with their corresponding regression
coefficient.  Visually, from the data scattering, it cannot be
expected that a simple empirical law can be found. Indeed, four
simple laws are indicated as not giving a nice regression
coefficient.
\newline
Note that these trial fits with simple empirical laws  for the  "all
Saints" case, i.e. Fig. \ref{fig9:Plot1NinhabitStrlili639} and Fig.
\ref{fig8:Plot1NinhabitStrlolo3folo},
are not nice enough to suggest a decomposition between males and females. 
\newline
It is also worth to note that  from  Fig.
\ref{fig8:Plot1NinhabitStrlolo3folo} it is confirmed that even if
regularities of power-law type does not apply for an entire set of
data, such regularity may exist for qualified subclusters. This is
precisely the case of the hagiotoponym Italian cities, for which
four "regimes" seem to emerge. One is for the 7 or 8  most frequent
(in population number) names. The next regime contains about 30
names, and the next one about 50. The final regime contains about
100 names. The first and third regimes have approximately the same
power law exponent.} \label{fig8:Plot1NinhabitStrlolo3folo}
\end{figure}

Zipf (1949) had observed  that a large number of  size
distributions, $N_r$ can  be approximated by a simple  {\it scaling
(power) law} $N_r = N_1/r $,   where  $r$ is the ranking parameter,
with  $N_{r } \ge N_{r+1}$, with obviously  $r<r+1$.
\newline
A more flexible equation, with two parameters,  reading
\begin{equation}\label{Zipfeq}
N_r = \cfrac{N_1}{r^{\alpha}},
\end{equation}
is called the rank-size scaling law and has been often applied to
city sizes. The particular case  ${\alpha}=1$ is thought to
represent a desirable situation, in which   forces of concentration
balance those of decentralization. Such a case is called  the
rank-size rule. The interested reader is referred to Gabaix (1999),
Gibrat (1957), Laherr\`ere and Sornette (1998), Ausloos (2013),
Fairthorne (1969), Adamic (2005), Shannon (1948), Vitanov and
Dimitrov (2014), McKean et al. (2009), Wolfe (2009, 2010), Lin
(2010), Wieder (2009). Hence, the rank-size relationship   has  been
frequently identified and sufficiently discussed to allow us  to
base much of the present investigation on such a simple law. This
may be "simply" because the rank-size relationship can be applied to
a wide range of specific situations (see Mart\`{\i}nez-Mekler et
al., 2009) and Zipf's law obtained in different models: one  example
is tied to the maximization of the entropy concept in Chen (2012);
another stems from  the law of proportionate effect, so called
Gibrat's law (Gibrat, 1957).
\newline
Thus, let us express Zipf's law, in other words:  the {\it
rank-frequency} relationship, i.e. the relationship between the
number $f$ (frequency) of the occurrence of an "event" and its rank
$r$ (Hill, 1974), consists of an inverse power law:

     \begin{equation}\label{V1}
     f\sim r^{-\alpha}.
       \end{equation}
A display of the rank-size (or rank-frequency, two names for the
same concept) relationship for  cities bearing a Saint name ranked
in decreasing order according to their "frequency",   Eq.
(\ref{V1}), is shown in Fig. \ref{Plot21fig3loloZM}. The best (least
square) corresponding power law fits are indicated; the gender (f or
m)  is distinguished beside the overall    (a)  size (s) data. It
can be observed that the power law fits look quite acceptable, being
almost perfect in the "female case".
\newline
Another law is attributed to Zipf : the  {\it size-frequency}
relationship, i.e.: the link between the frequency $f$ and the size
$s$ of an "event", is also in this case an inverse power law:

     \begin{equation}\label{V2}
     f\sim s^{-\lambda}.
       \end{equation}
Of course, deviations from the simple law  often occur, as
illustrated through Fig. \ref{fig8:Plot1NinhabitStrlolo3folo} and
Fig. \ref{fig8b:Plot6ATI5lolo639fits4}.   In fact, there is no
obligation for the size-frequency data to be enveloped by a purely
convex or purely concave function (Egghe and Waltman, 2011). A so
called "king effect" (Lah\`errere and Sornette, 1998) i.e. a sharp
upturn at low $r$ values often exists. A leveling off at low $r$,
the queen effect can also occurs, as in   Fig.
\ref{Plot21fig3loloZM} (see also Ausloos, 2013). In such a case, a
Zipf-Mandelbrot-like (ZM), sometimes called
Bradford-Zipf-Mandelbrot-like (BZM), law

   \begin{equation}\label{ZMlikeCr}
  J(r) =\frac{J^{*}}{(\nu+r)^\zeta},
  \end{equation}
might be considered as more realistic (Fairthorne, 1969).  It
implies three parameters ($J^*$, $\nu$ and $\zeta$).
\newline
Moreover, it  can also be asked how many times one can find an
"event" greater than some size $s$, i.e.  within the size-frequency
relationship. Pareto (1896) found out that  the the cumulative
distribution function of such events  follows an inverse power of
$s$, or  in other words,

    \begin{equation}\label{V3}P\;[Y>s] \sim s^{-\kappa},\end{equation}
where $Y$ is the random variable of the size, $P$ a probability
measure and $\kappa$ a scalar. Again, this is quite an
approximation, as illustrated through Fig. \ref{fig3:Plot41CDFlolo}.
\newline
It is important to observe that the Yule-Simon distribution can be
used to reproduce a Zipf law, but it introduces an exponential
cutoff in the upper tail (Rose et al., 2002). The stretched
exponentials
 (Lah\`errere and Sornette, 1998) and log-normal
distributions (Montroll and Shlesinger, 1983) usually reproduce one
of the tails but not the other. Usually, such deviations do not
change in a dramatic way the correlation coefficient  since the
tails do not have a great impact upon this coefficient. Moreover,
such distributions assume along (infinite) tail  which is
antagonistic to the concept of  finite sizes, when there is a true
maximum rank $r_M$.
\newline
When an inflection point occurs, the  2-parameter form, so called
Lavalette function (Lavalette, 1966),  of the  rank-frequency (or
rank-size) relationship reads:
 \begin{equation} \label{Lavalette2}
g_2(r)= \kappa_2\; \left[\frac{N\;r}{ N-r+1} \right] ^{-\chi};
\end{equation}
its simple generalization into a 3-parameter free function (Popescu
et a., 1997, Popescu, 2003, Mansilia et al., 2007, Ausloos, 2014b)
is
  \begin{equation} \label{Lavalette3a}
 \;\;  g_3(r)= \kappa_3\;  \frac{(N\;r)^{- \gamma}}  { (N-r+1)^{-\xi}  }.
\end{equation}
Note also that the role of $r$ as independent variable in Zipf's
law. Specifically, Eq. (\ref{Zipfeq}) is taken by the ratio $r/(N -
r + 1)$ between the descending and the ascending ranking numbers.
The semi-logarithmic graph shows a reverse sigmoidal S-shape (or an
inverse N-shape) which cannot be provided by  Zipf's law.  By the
way, in a double-logarithmic diagram the downwards deviation from
the Zipf's straight line at high rank is much emphasized.
\newline
More complicated forms with many more  parameters generalize the
Lavalette form  (see Ausloos, 2014a, 2014b and Voloshynovska, 2011).
They provide better fits, but seem of no special interest here.
\newline
It is fair to mention that the mere hyperbolic form has some  sound
mathematical, statistical and physical basis. The  BZM form,
Eq.(\ref{ZMlikeCr}), and the Lavalette functions  have not yet
received a sound physical basis to our knowledge though some
mathematical insight has been provided (Naumis and Cocho, 2008).
\subsection*{Kendall $\tau$ coefficient}
The Kendall's $\tau$ measure, introduced in Kendall (1938) compares
the number of concordant pairs $p$ and non-concordant pairs $q$
through
\begin{equation}\label{taueq}
\tau = \; \frac{p-q}{p+q}.
\end{equation}
Of course, $p+q= N(N-1)/2$, where $N$ is the number of  measures,
when there is no  doubt about measure rank (i.e., no rank overlap).
For large samples, it is also common to measure
 \begin{equation}\label{tauvar}
Z=\frac{\tau}{\sigma_{\tau}}\;\equiv
\frac{\tau}{\sqrt{\frac{2(2N+5)}{9N(N-1)}}}.
\end{equation}
similar to the classical one,  when the distribution can be
approximated by  the normal distribution, with mean zero and
variance,  - in order to emphasize the coefficient $\tau$
significance. From  a purely statistical perspective, under the null
hypothesis of independence of the rank sets,  such a sampling would
have an expected value $\tau$ and $Z$  = 0.

A website (Wessa, 2012) allows $\tau$ immediate calculation.


\newpage
\section*{Appendix B: Figures and Tables}
\subsection*{Tables}

\begin{table}
  \begin{center} \begin{tabular}{|c|c|c|c|}
\hline
Municipality &   Saint name & Province& Region     \\ \hline
Guardia Sanframondi & FREMONDO &BN &CAMPANIA \\
Sampeyre & PIETRO &CN &PIEMONTE \\
Samugheo & MICHELE &OR & SARDEGNA\\
Sanfr\'{e} & IFFREDO & CN&PIEMONTE \\
Sanfront  & FRONTONE &CN &PIEMONTE \\
Sangiano  & GIOVANNI& VA&LOMBARDIA \\
Sanremo & ROMOLO &IM &LIGURIA \\
Santeramo in Colle & ERASMO &BA &PUGLIA \\
Santhi\`a  & \textit{AGATA} & VC& PIEMONTE\\
Santomenna & MENNA &SA &CAMPANIA \\
Santorso & ORSO & VI&VENETO \\
Santu Lussurgiu & LUSSORIO &OR &SARDEGNA \\
Sanzeno & SISINNIO & TN & TRENTINO ALTO ADIGE\\
 \hline \end{tabular}  \end{center}
 \caption{So called accepted strange cases: municipality  (listed in alphabetical order) and corresponding
Saint name, with the province or region membership. The (only)
female Saint is  reported \textit{in italic}. Very interesting are
the cases of \textit{Sanremo} in Liguria and \textit{Sanzeno} in
Trentino Alto Adige. Unexpectedly Sanremo, which reasonably should
refer to Saint Remo, points actually to Saint Romolo, bishop of
Genua in the IX century. Indeed, a Saint named Remo does not exist,
and Sanremo is a dialectal contraction of San R{\oe}m\"{u}, which
means Romolo. Sanzeno who might  let the reader think  about  Saint
Zenone derives actually from the martyr
Sisinnio.}\label{tab:strange-cases}
\end{table}

\begin{table}
  \begin{center} \begin{tabular}{|c|c|c|c|c|c|c|}
\hline REGION  &N. Cities &Hagiotop.  &   Saint  freq. &  Males   &
Females \\ \hline
ABRUZZO &   305 &   24  &   24  &   18  &   6   \\
BASILICATA  &   131 &   11  &   11  &   11  &   0   \\
CALABRIA    &   409 &   59  &   59  &   48  &   11  \\
CAMPANIA    &   551 &   68  &   68  &   59  &   9   \\
EMILIA ROMAGNA  &   348 &   29  &   29  &   25  &   4   \\
FRIULI VENEZIA GIULIA   &   218 &   17  &   17  &   16  &   1   \\
LAZIO   &   378 &   31(*)   &   32  &   30  &   2   \\
LIGURIA &   235 &   13  &   13  &   12  &   1   \\
LOMBARDIA   &   1544    &   75  &   75  &   67  &   8   \\
MARCHE  &   239 &   27  &   27  &   25  &   2   \\
MOLISE  &   136 &   15  &   15  &   13  &   2   \\
PIEMONTE   &   1206    &   63  &   63  &   59  &   4   \\
PUGLIA  &   258 &   28(*)   &   29  &   26  &   3   \\
SARDEGNA    &   377 &   26  &   26  &   20  &   6   \\
SICILIA &   390 &   40  &   40  &   25  &   15  \\
TOSCANA &   287 &   20  &   20  &   17  &   3   \\
TRENTINO ALTO ADIGE &   333 &   16  &   16  &   14  &   2   \\
UMBRIA  &   92  &   5   &   5   &   3   &   2   \\
VALLE D'AOSTA   &   74  &   16  &   16  &   15  &   1   \\
VENETO  &   581 &   54  &   54  &   45  &   9   \\
\hline Total  &   8092&637    &   639 &   548 &   91\\ \hline
                                    \end{tabular}  \end{center}
\caption{Summary Table with the hagiotoponym cities in IT. Data are
disaggregated at the  regional level, in the alphabetical order for
regions. The discrepancy between the number of hagiotoponyms (637)
and the Saints "frequency" (639) is due to the presence of cities
with a name containing two Saints: Cosma and Damiano, Marzano and
Giuseppe, see $(^*)$ in the Table and the discussion in the text.
All data refer to 2011.} \label{tab:Tabla1}
\end{table}

\begin{table}
  \begin{center} \begin{tabular}{|c|c|}
\hline
Group of linguistic transformation &   Saint name      \\
\hline\hline
Basile, Basilio & BASILE \\
Biagio, Biase & BIAGIO \\
Casciano, Cassiano & CASCIANO \\
Cesario, Cesareo & CESARIO \\
Cosma, Cosmo & COSMO \\
Donato, Don\'a & DONATO\\
Fedele, Fele & FEDELE \\
Felice, Fili & FELICE \\
Floriano, Fiorano & FLORIANO \\
Floro, Fior & FLORO \\
Gemini & GEMINE \\
Genesio, Ginesio & GENESIO \\
Lorenzo, Lorenzello & LORENZO\\
Maria, Marie, Madonna, Notre-Dame & \textit{MARIA} \\
Michele, Sammichele & MICHELE \\
Nazzaro, Nazario, Sannazzaro & NAZZARO \\
Nicandro, Sannicandro & NICANDRO\\
Nicola, Niccol\'{o}, Nicolao, Nicol\'{o}, Sannicola & NICOLA \\
Paolo, Polo, sampolo & PAOLO \\
Pietro, Piero, Pier, sampiero, Pietrangeli & PIETRO \\
Quirico, Chirico & QUIRICO \\
Stefano, Stino & STEFANO\\
Zenone, Zeno & ZENONE \\
\hline
\end{tabular}  \end{center}
\caption{Linguistic transformations and corresponding Saint name.
The only female Saint name so concerned  is Maria, as emphasized
\textit{in italic}.} \label{tab:ling-trans}
\end{table}

\begin{table}
  \begin{center} \begin{tabular}{|c|c|}
\hline
Municipality &   Saint name  \\
\hline\hline
San Candido/Innichen & CANDIDO \\
Santa Cristina Valgardena/St. Christina in Gr{\"o}den & \textit{CRISTINA} \\
Senale-San Felice/Unsere Liebe Frau im Walde-St. Felix
& FELICE \\
San Genesio Atesino/Jenesien & GENESIO \\
San Leonardo in Passiria/St. Leonhard in Passeier & LEONARDO \\
San Lorenzo di Sebato/St. Lorenzen  & LORENZO \\
San Martino in Badia/St. Martin in Thurn & MARTINO \\
San Martino in Passiria/St. Martin in Passeier & MARTINO \\
San Pancrazio/St. Pankraz & PANCRAZIO \\
\hline
Antey-Saint-Andr{\' e} & ANDREA \\
Challand-Saint-Anselme & ANSELMO \\
Saint-Christophe & CRISTOFORO\\
Saint-Denis & DENIS \\
Pr{\' e}-Saint-Didier & DIDERO \\
Rh{\^e}mes-Saint-Georges & GIORGIO\\
Gressoney-Saint-Jean & GIOVANNI \\
Saint-Marcel & MARCELLO \\
Rh{\^e}mes-Notre-Dame & \textit{MARIA} \\
Pont-Saint-Martin & MARTINO \\
Saint-Nicolas & NICOLA\\
Saint-Oyen & OYEN\\
Saint-Pierre & PIETRO \\
Saint-Rh{\' e}my-en-Bosses & RHEMY \\
Saint-Vincent & VINCENZO\\
Challand-Saint-Victor & VITTORIO \\
\hline
\end{tabular}  \end{center}
\caption{The set of  9+16 toponyms in foreign (= non-Italian)
language and the corresponding Saint name (in Italian). They belong
to Valle d'Aosta (16 cities) and Trentino Alto Adige (9 cities).
Female Saints are reported \textit{in italic}.}
\label{tab:strangers}
\end{table}


    \begin{table}
  \begin{center} \begin{tabular}{|c|c|c|c|c|c|}
\hline SAINT NAME  &   Freq.   &   SAINT NAME  &   Freq.   &   SAINT
NAME  &   Freq. \\ \hline
PIETRO      &   43  &      $CRISTINA$   &   4   &      $ANASTASIA$  &   2   \\
    &       &      $MARGHERITA$     &   4   &      $DOMENICA$   &   2   \\
GIOVANNI        &   36  &           &       &      $EUFEMIA$    &   2   \\
    &       &      BARTOLOMEO   &   4   &      $GIUSTINA$   &   2   \\
MARTINO     &   27  &      CASCIANO     &   4   &      $MARINA$     &   2   \\
GIORGIO     &   27  &      \underline{DAMIANO } &   \underline{4}   &      $SOFIA$  &   2   \\
        &       &      GENESIO  &   4   &      $TERESA$     &   2   \\
$MARIA$     &   23  &      GERMANO  &   4   &           &       \\
        &       &      GIULIANO     &   4   &      CARLO    &   2   \\
STEFANO     &   19  &      \underline{GIUSEPPE }    &   \underline{4}   &      \underline{COSMA  }  &   \underline{2}   \\
    &       &      MARCELLO     &   4   &      CONSTATINO   &   2   \\
NICOLA      &   16  &           &       &      COSTANZO     &   2   \\
PAOLO       &   16  &      $ANNA$   &   3   &      CRISTOFORO   &   2   \\
    &       &      $CATERINA$   &   3   &      DANIELE  &   2   \\
LORENZO     &   14  &      $ELENA$  &   3   &      DEMETRIO     &   2   \\
VITO        &   14  &      $VITTORIA$   &   3   &      DIDERO   &   2   \\
        &       &           &       &      EGIDIO   &   2   \\
$AGATA$     &   12  &      ALESSIO  &   3   &      ELPIDIO  &   2   \\
        &       &      AMBROGIO     &   3   &      EUSANIO  &   2   \\
MICHELE     &   11  &      BASILE   &   3   &      FEDELE   &   2   \\
    &       &      CESARIO  &   3   &      FERDINANDO   &   2   \\
ANDREA      &   9   &      CIPRIANO     &   3   &      FERMO    &   2   \\
GIACOMO     &   9   &      COLOMBANO    &   3   &      FLORIANO     &   2   \\
MARCO       &   9   &      ELIA     &   3   &      FLORO    &   2   \\
        &       &      GERVASIO     &   3   &      GIUSTO   &   2   \\
$LUCIA$     &    7  &      MANGO    &   3   &      ILARIO   &   2   \\
        &       &      \underline{MARZANO}  &   \underline{3}   &      LEONARDO     &   2   \\
BENEDETTO       &   8   &      ROCCO    &   3   &      MAURIZIO     &   2   \\
FELICE      &   8   &      SEBASTIANO   &   3   &      NICANDRO     &   2   \\
MAURO       &   8   &      SECONDO  &   3   &      PANCRAZIO    &   2   \\
    &       &      SEVERINO     &   3   &      POTITO   &   2   \\
ANTONIO     &   6   &           &       &      SIRO     &   2   \\
BIAGIO      &   6   &           &       &      TEODORO  &   2   \\
GREGORIO        &   6   &           &       &      URBANO   &   2   \\
    &       &           &       &      VALENTINO    &   2   \\
DONATO      &   5   &           &       &      VITTORE  &   2   \\
NAZZARO     &   5   &       &       &       &       \\
QUIRICO     &   5   &       &       &       &       \\
VINCENZO        &   5   &       &       &       &       \\
ZENONE      &   5   &       &       &       &       \\
   \hline
\end{tabular}  \end{center}
\caption{Different Saints names in alphabetical order with frequency
(Freq.), or "popularity", when occurring more than once in
hagiotoponyms;  distinguishing males  (74) from females (16) whose
names are italicized. Recall that the underlined Saints should count
for 1/2 when counting the number of cities.} \label{tab:Tabla2}
\end{table}

\begin{table}
  \begin{center} \begin{tabular}{|c|c|c|c|}
\hline
\textit{ANATOLIA}    &   ABBONDIO    &   FIDENZIO    &   ORSO    \\
\textit{BRIGIDA} &   AGAPITO &   FILIPPO &   OYEN    \\
\textit{CESAREA} &   AGNELLO &   FRANCESCO   &   PATRIZIO    \\

\textit{ELISABETTA}   &   AGOSTINO    &   FRATELLO    &   PELLEGRINO  \\

\textit{FIORA}  &   ALBANO  &   FREMONDO    &  PIO  \\

\textit{FLAVIA}   &   ALESSANDRO  &   FRONTONE    &   PONSO \\

\textit{GIULETTA}  &   ALFIO   &   GAVINO  &   POSSIDONIO   \\

\textit{GIUSTA}   &   ANSELMO &   GEMINE  &   PRISCO  \\

\textit{MARINELLA}   &   ANTIMO  &   GENNARO &   PROCOPIO  \\

\textit{NINFA} &   ANTIOCO &   GILLIO  &   PROSPERO    \\

\textit{ORSOLA}  &   ANTONINO    &   GIMIGNANO   &   QUIRINO    \\

\textit{PAOLINA} &   APOLLINARE  &   GIULIO  &   RAFFAELE \\

\textit{SEVERINA} &   ARPINO  &   GIUSTINO    &   RHEMY    \\

\textit{SUSANNA}&   ARSENIO &   GODENZO &   ROBERTO    \\

\textit{VENERINA}
    &   BASSANO &   IFFREDO &   ROMANO   \\
    &   BELLINO &   IPPOLITO    &   ROMOLO \\
    &   BENIGNO &   LAZZARO &   RUFO  \\
    &   BERNARDINO  &   LEO &   SAVINO   \\
    &   BONIFACIO   &   LEUCIO  &   SEVERO  \\
    &   BOVO    &   LUCA    &   SISINNIO  \\
    &   BRUNO   &   LUCIDO  &   SOSSIO    \\
    &   CALOGERO    &   LUPO    &   SOSTENE  \\
    &   CANDIDO &   LUSSORIO    &   SOSTI \\
    &   CANZIAN &   MAGNO   &   SPERATE   \\
    &   CATALDO &   MAMETTE &   TAMMARO \\
    &   CIPIRELLO   &   MARCELLINO  &   TOMASO \\
    &   CLEMENTE    &   MASSIMO &   VENANZO  \\
    &   CONO    &   MENNA   &   VENDEMIANO \\
    &   DALMAZZO    &   MINIATO &   VERO  \\
    &   DENIS   &   OLCESE  &   VICINO    \\
    &   DOMENICO    &   OMERO   &   VITALE  \\
    &   DONACI  &   OMOBONO &   VITALIANO  \\
    &   DORLIGO &   ONOFRIO &   VITTORIO   \\
    &   ERASMO  &   ORESTE  &        \\ \hline

\end{tabular}  \end{center}
\caption{Hapax Saints: first column collects the 15 females
(\textit{in italic}), while the other columns list the 101 males.}
\label{tab:Tabla5}
\end{table}

\begin{table}
\begin{center} \begin{tabular}{|c|c|c|c|c|c|}
\hline St. Name    &   Sex &   Toponym &   REGION  &   N.  inhab. \\
\hline

Giovanni   & M &  Sesto San Giovanni & LOMBARDIA  &      76970
\\
\textit{Elena} & F &Quartu Sant'Elena  & SARDEGNA  &    69295 \\
\underline{Severo} &M  & San Severo &PUGLIA  &     55053 \\
\underline{Romolo}  &  M & Sanremo & LIGURIA & 53617 \\
Benedetto &M &  San Benedetto del Tronto  & MARCHE &46988 \\
Giorgio &M &San Giorgio a Cremano & CAMPANIA   & 45058\\
Donato & M &San Don\'{a} di Piave & VENETO & 40691\\
Giuliano & M &San Giuliano Milanese &LOMBARDIA & 35924\\
\underline{Antimo} & M &Sant'Antimo &CAMPANIA &33950 \\
\textit{Maria} &F &Santa Maria Capua Vetere  & CAMPANIA & 32603\\
\underline{Lazzaro} &M &San Lazzaro di Savena &EMILIA ROMAGNA &31 183\\
Giuliano &   M  & San Giuliano Terme & TOSCANA &     31 157 \\
Donato &M &San Donato Milanese &LOMBARDIA  &      31 037 \\
\underline{Miniato} &M & San Miniato &TOSCANA &     27633\\
Giovanni &   M  & San Giovanni Rotondo & PUGLIA & 27371 \\
Giuseppe & M & San Giuseppe Vesuviano &CAMPANIA & 27310 \\
Giovanni & M & San Giovanni in Persiceto  & EMILIA ROMAGNA & 27051
\\
Erasmo & M & Santeramo in Colle  &PUGLIA &     26662 \\
Elpidio &M &Porto Sant'Elpidio  &MARCHE &     25354 \\
\textit{Anastasia}  & F & Sant'Anastasia &CAMPANIA  &       25082 \\
... &   ... &   ... &   ... &   ...  \\ \hline
\end{tabular}  \end{center}
\caption{Top 20 hagiotoponym cities in IT, in terms of number of
inhabitants. Females are italicized; hapax Saints  are underlined.}
\label{tab:Tabla3}
\end{table}

    \begin{table}
  \begin{center} \begin{tabular}{|c|c|c|c|c|c|}
\hline St. Name    &   Sex &   Toponym &   REGION  &   N.  inhab. \\
\hline

... &   ... &   ... &   ... &   ...  \\
Paolo  &  M  &  San Paolo Albanese &  BASILICATA  &      313 \\
\textit{Eufemia} & F & Sant'Eufemia a Maiella &  ABRUZZO  &     304 \\
\underline{Vicino} & M & Poggio San Vicino  &  MARCHE  &      297 \\
\underline{Ponso}  &  M & San Ponso & PIEMONTE & 279 \\
\textit{Elena} & F  &  Sant'Elena Sannita &  MOLISE & 272 \\
Michele & M & Olivetta San Michele  & LIGURIA  &     225 \\
\underline{Oyen} & M & Saint-Oyen &  VALLE D'AOSTA & 217 \\
Giovanni & M & San Giovanni Lipioni &   ABRUZZO & 213 \\
Biagio &  M & San Biase & MOLISE & 209 \\
Giorgio & M & Rhemes-Saint-Georges & VALLE D'AOSTA & 196 \\
Benedetto & M  & San Benedetto Belbo & PIEMONTE & 193 \\
Giovanni & M & Sale San Giovanni & PIEMONTE  & 178 \\
Vito & M & Celle di San Vito & PUGLIA & 172 \\
\textit{Lucia} & F &   Villa Santa Lucia degli Abruzzi & ABRUZZO & 144 \\
Paolo & M & San Paolo Cervo & PIEMONTE & 142 \\
Giorgio & M & San Giorgio Scarampi & PIEMONTE & 131 \\
Benedetto & M & San Benedetto in Perillis & ABRUZZO & 130 \\
\textit{Maria} & F & Rhemes-Notre-Dame & VALLE D'AOSTA & 114 \\
Stefano & M & Santo Stefano di Sessanio  & ABRUZZO & 114 \\
Giuseppe & M & Rima San Giuseppe & PIEMONTE & 68 \\
\hline
\end{tabular}  \end{center}
\caption{Bottom 20 hagiotoponym cities in IT, in terms of the number
of inhabitants. Females are italicized; hapax Saints are
underlined.}\label{tab:Tabla4}
\end{table}

\begin{table}
\begin{center} \begin{tabular}{|c|c|c|c|c|c|}
\hline Name    &   Sex &   Toponym &   REGION  &   Average ATI
\\ \hline

Giovanni  &  M&  Sesto San Giovanni &LOMBARDIA&  1240078983
\\
\textit{Elena}& F&  Quartu Sant'Elena&  SARDEGNA&   723839629\\
Donato& M& San Donato Milanese&LOMBARDIA&  681177081\\
\underline{Romolo}&   M & Sanremo& LIGURIA&650834136\\
\underline{Lazzaro}&M&  San Lazzaro di Savena & EMILIA ROMAGNA& 589480816\\
Benedetto&  M&  San Benedetto del Tronto&MARCHE & 558246404\\
Donato& M&  San Donà di Piave & VENETO & 543509572\\
Giuliano& M& San Giuliano Milanese& LOMBARDIA&507215120 \\
Giuliano & M&  San Giuliano Terme& TOSCANA&458205438\\
Giorgio&M & San Giorgio a Cremano& CAMPANIA& 427785711
\\
Giovanni&M& San Giovanni in Persiceto& EMILIA ROMAGNA& 396583755\\
\underline{Miniato}&M& San Miniato & TOSCANA&350427626 \\
\underline{Severo}& M& San Severo& PUGLIA & 346136016 \\
Pietro& M&  Castel San Pietro Terme&EMILIA ROMAGNA & 316883099\\
Giovanni&   M&  San Giovanni Lupatoto&  VENETO & 312392144
\\
\textit{Maria} & F& Santa Maria Capua Vetere& CAMPANIA & 304644801 \\
Mauro& M& San Mauro Torinese &PIEMONTE & 302387316\\
Elpidio & M& Porto Sant'Elpidio &MARCHE &247343572 \\
Casciano&   M &San Casciano in Val di Pesa&TOSCANA & 238256972\\
Bonifacio&  M&  San Bonifacio& VENETO &236053538\\
... &   ... &   ... &   ... &   ...  \\ \hline
\end{tabular}  \end{center}
\caption{Top 20 hagiotoponym cities in IT, in terms of the $<ATI>$.
Females are italicized; hapax Saints are
underlined.}\label{tab:Tabla3-ATI}
\end{table}

    \begin{table}
  \begin{center} \begin{tabular}{|c|c|c|c|c|c|}
\hline Name    &   Sex &   Toponym &   REGION  &   Average ATI
\\ \hline
... &   ... &   ... &   ... &   ...  \\
\textit{Eufemia} &F&  Sant'Eufemia a Maiella& ABRUZZO&2267311 \\
Giovanni& M & Sale San Giovanni & PIEMONTE&   2227025\\
Pietro& M& San Pietro in Amantea&  CALABRIA&   2219842\\
\underline{Menna}&  M & Santomenna& CAMPANIA & 2076634\\
Michele&M & Olivetta San Michele & LIGURIA&2070318\\
Benedetto & M & San Benedetto Belbo&PIEMONTE & 2019683\\
Paolo & M & San Paolo Cervo&PIEMONTE & 2006905 \\
Biagio& M& San Biagio Saracinisco& LAZIO & 1992708 \\
Alessio&M & Sant'Alessio in Aspromonte& CALABRIA & 1879383\\
Giovanni &  M  & San Giovanni Lipioni&   ABRUZZO & 1802324\\
Nazzaro&M&  San Nazzaro Val Cavargna & LOMBARDIA&  1603334
\\
\textit{Elena} & F& Sant'Elena Sannita &MOLISE & 1537497\\
\textit{Maria} &  F & Rhemes-Notre-Dame & VALLE D'AOSTA & 1458916\\
Vito & M & Celle di San Vito & PUGLIA& 1309750\\
Biagio&M & San Biase & MOLISE &1199419 \\
Benedetto&  M & San Benedetto in Perillis& ABRUZZO&1191932\\
\textit{Lucia} & F & Villa Santa Lucia degli Abruzzi&ABRUZZO & 982732\\
Stefano&M & Santo Stefano di Sessanio & ABRUZZO&970598 \\
Giuseppe & M & Rima San Giuseppe & PIEMONTE & 902302 \\
Giorgio&M & San Giorgio Scarampi & PIEMONTE & 775239 \\
\hline
\end{tabular}  \end{center}
\caption{Bottom 20  hagiotoponym cities in IT, in terms of $<ATI>$.
Females are italicized.   The only hapax Saint in this Table is
underlined.}\label{tab:Tabla4-ATI}
\end{table}

\begin{table} \begin{center}
\begin{tabular}[t]{|c|c|c|c|}
  \hline
Statistical indicator& Name  popularity  & Population &Average ATI \\
\hline
Minimum &   1 &     68   &  7.752 e+05  \\
Maximum &   43 &    7.697 e+04   &  1.240 e+09  \\
Sum &   639 &   3.391 e+06   &  3.486 e+10  \\
N. data points  &   206  &  639  &  639 \\
Mean    &   3.102    &  5.306 e+03&     5.455  e+07 \\
Median  &   1  &    2.667 e+03     &  2.366 e+07  \\
RMS &   6.215    &  9.452 e+03&     1.108 e+08  \\
Std. Deviation  &   5.399    &  7.829 e+03   &  9.647 e+07  \\
Variance    &   29.146   &  6.130 e+07   &  9.307 e+15  \\
Std Error   &   0.3761   &  309.72   &  3.816 e+06  \\
Skewness    &   4.550    &  4.2325   &  5.699   \\
Kurtosis    &   24.007   &  25.347   &  47.906  \\ \hline
Mean / Std. Dev &0.5746& 0.6777 &0.5654 \\
3(Mean-Median)/Var. &0.2160 & 1.29 e-04& 9.96 e-09\\\hline
\end{tabular}
\caption{Summary of distribution statistical characteristics for the
Saint name popularity, i.e. measured by the number of hagiotoponym
cities with an equivalent Saint name, for the number of inhabitants
and for the $<ATI>$ of Italian
  hagiotoponym cities. 
   }\label{TableATIstat}
\end{center} \end{table}

\begin{table} \begin{center}
\begin{tabular}[t]{cccccccc}
  \hline
   $ $   & Population  & Average ATI \\
\hline
Minimum  & 30 &3.3219 e+05  \\
Maximum &2.6637 e+06 &4.4726 e+10 \\
Sum & 5.9570 e+07 &7.0738 e+11    \\
N. data points  &8092 &8092   \\
Mean ($\mu$) &  7361.6635 &8.7417 e+07 \\
Median ($m$) & 2443 &2.3828 e+07 \\
RMS & 40927.3114 &6.682 e+08 \\
Std. Deviation ($\sigma$) & 40262.2783 &6.6256 e+08\\
Variance & 1.6210 e+09 &4.3899 e+17 \\
Std. Error & 447.5797 &7.3654 e+06 \\
Skewness  & 43.7288 &49.126 \\
Kurtosis  & 2545.1474 &2955.2    \\  \hline
 $\mu/\sigma$  & 0.1828 & 0.1319  \\
$3(\mu-m)/\sigma$  & 9.1027 e-06 & 0.2879   \\
 \hline
 \end{tabular}
\caption{Summary of  (rounded) statistical characteristics for the
the number of inhabitants in 2011 and the average ATI (in EUR) over
the quinquennium 2007-2011 of IT cities
($N=8092$).}\label{Tablestat-IT}
 \end{center} \end{table}

\begin{table}
\begin{center} \begin{tabular}{|c|c|c|c|c|c|}
   \hline& &&&  \\ Statistical indicator & $ATI(x)$ & $POP(x)$ & $\overline{ATI}(x)$
& $\overline{POP}(x)$ \\ \hline
Minimum &   2.0766 e+06 &   217 &   2.0766 e+06 &   217     \\
Maximum &   4.0187 e+09 &   3.4461 e+05 &   6.5083 e+08 &   55053   \\
Sum &   3.4857 e+10 &   3.3906 e+06 &   1.1433 e+10 &   1.1776 e+06 \\
Mean    &   1.6921 e+08 &   16459   &   5.5500 e+07 &   5716.5  \\
Median  &   6.0355 e+07 &   5885.5  &   3.4461 e+07 &   3524.5  \\
RMS &   4.2219 e+08 &   37802   &   9.6161 e+07 &   9263.7  \\
Std. Deviation  &   3.8774 e+08 &   34114   &   7.8720 e+07 &   7307.3  \\
Variance    &   1.5035 e+17 &   1.1638 e+09 &   6.1968 e+15 &   5.3397 e+07 \\
Std. Error   &   2.7015 e+07 &   2376.8  &   5.4847 e+06 &   509.12  \\
Skewness    &   6.262   &   5.879   &   4.532   &   3.886   \\
Kurtosis    &   50.840  &   45.537  &   26.310  &   19.521  \\
\hline
Mean/ Std. Dev. &  0.4364  &  0.4825&  0.7050  &  0.7823  \\
3(Mean-Median)/ Var. &  2.172 e-11  &  2.726 e-05 &   3.395  e-09&  4.105  e-05 \\
\hline
\end{tabular}  \end{center}
\caption{Statistical indicators for the treated  (206  different
Saints) data $ATI(x)$, $POP(x)$, $\overline{ATI}(x)$ and
$\overline{POP}(x)$. 
}\label{tab:stat-treated-data}
\end{table}

\begin{table} \begin{center}
\begin{tabular}[t]{cccc}
  \hline
&(i) &  (ii) \\ \hline
Kendall $\tau$& 0.849  \\ 
$p+q$&32 736 186 \\
 $  p-q$    &27 778 116  \\
$p$&30 256 042 \\
$q$&2 480 144 \\
$Z$ (Eq. (\ref{tauvar}))& 114.6 \\
 \hline
\end{tabular}
\caption{Kendall $\tau$, Eq. (\ref{taueq}), correlation statistics
of ranking order between the Number of inhabitants in 8092 cities
and the corresponding averaged ATI over the period $2007-2011$.
 } \label{Tabletaurank-IT}
\end{center} \end{table}

\begin{table} \begin{center}
\begin{tabular}[t]{c|cc|ccccc}
  \hline
rank correlation &(i) &  (ii) & (iii)   &(iv)       &   (v) &(vi)   
\\
between&$ POP(x)$  &$\overline{POP}(x)$     &  $POP(x)$   &$ATI(x)$   &$\overline{POP}(x)$    &     $\overline{ATI}(x)$    \\
and &  $ATI(x)$&      $\overline{ATI}(x)$&    $F$&   $F$&    $F$&
$F$
 \\ \hline \hline

Kendall $\tau$ 
&0.850& 0.788   &   0.510   &   0.518   &   0.068 & 0.102 \\
$p-q$   &17950& 16641  &   8660   &   8794      &   1169 &   1724 \\

$p+q$   &21114 &    21113  &   16964 &   16964 &   16964 &   16964 \\

$p$ &19532 &    18877  &   12812  &   12879    &   9066 &   9344 \\
$q$ &1582 & 2236   &   4152  &   4085      &   7897 &   7620 \\
 $Z=\tau/\sigma_{\tau}$ &18.145 &16.821 &   10.887  &   11.058  &   1.452  &   2.177 \\
 2-sided p-value  & 0.0000  & 0.0000  &  0.0000  & 0.0000&
0.17995 &  0.047265 \\
 \hline
\end{tabular}
\caption{Kendall $\tau$, Eq. (\ref{taueq}),  correlation statistics
of ranking order   between    (i) $ POP(x)$  and    $ATI(x)$ (ii)
$\overline{POP}(x)$ and    $\overline{ATI}(x)$;  (iii)  $POP(x)$ and
$F$; (iv)  $ATI(x)$  and $F$; (v)  $\overline{POP}(x)$  and $F$;
(vi)       $\overline{ATI}(x)$  and $F$, for the 206  Saints ($x$),
with notations as in the text; $\sigma_{\tau}$= 0.046844.}
 \label{TabletaurankATIPOPF2ndtype}
\end{center} \end{table}

\newpage
\subsection*{Figures}\label{sec:listoffigures}

      \begin{figure}
  \includegraphics[width=5.8in]
  {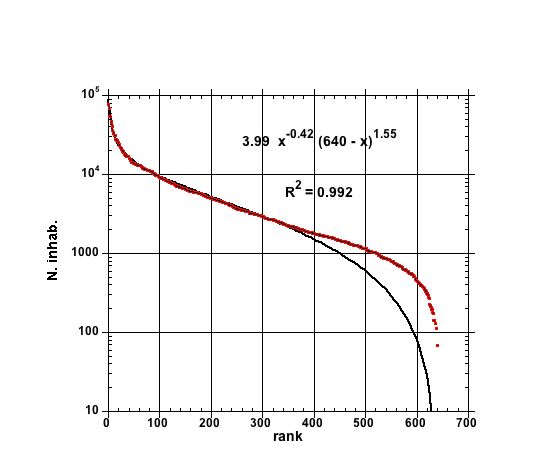} 
\caption{Semi-log display of the Number of inhabitants in the
 639   (637 + 2; see text) Saints generating Italian
hagiotoponyms,  -  cities ranked in decreasing order of the number
of inhabitants;   a fit by a Lavalette function shows a convincing
fit for $r\le350$.} \label{figA1:Plot3Nihab639lilo4Lav3}
\end{figure}

      \begin{figure}
  \includegraphics[width=5.8in]
  {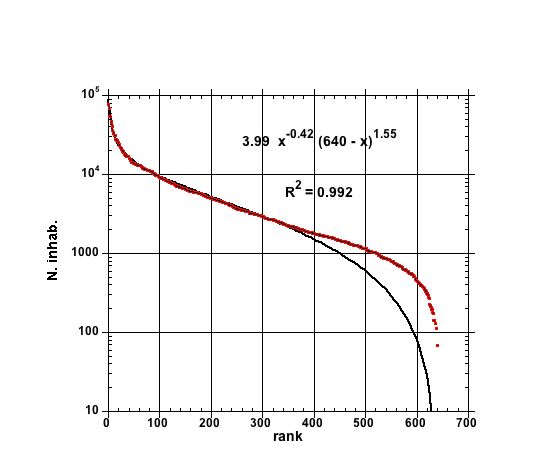} 
\caption{Histogram of the number of cities having a number of
inhabitants in some range. Big cities are emphasized, pointing to
the presence of outliers.} \label{Plot 1histoNinhab}
\end{figure}

      \begin{figure}
  \includegraphics[width=5.8in]
  {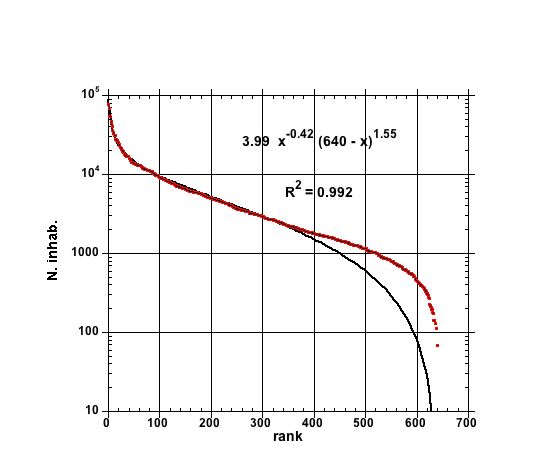} 
\caption{Log-log plot of the number of inhabitants vs. rank. $K$
represents the 8092 cities, while $C$ is the set of the upper
8092-80 cities, thus removing 80 outliers. It seems that the Lav3
function (green) fits the $C$ data very well; in contrast the fit
for the 8092 is not so good.} \label{Plot 4Lav3-80lolo}
\end{figure}

      \begin{figure}
  \includegraphics[width=5.8in]
  {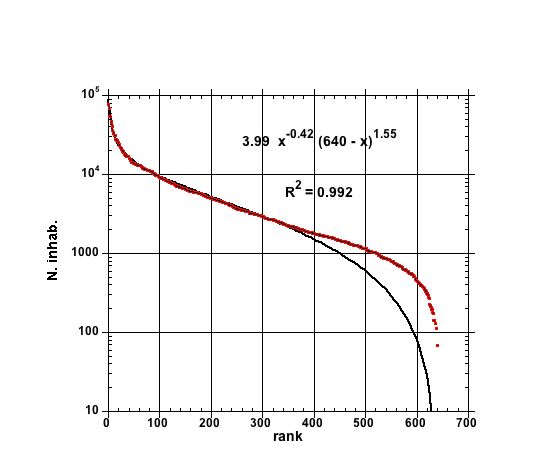} 
\caption{Also in this case, $K$ represents the 8092 cities, while
$C$ is the set of the upper 8092-80 cities, thus removing 80
outliers. As in Fig. \ref{Plot4Lav3-80lilo}, one barely sees that
the Lav3 function (green) fits the $C$ data very well; in contrast
the fit for the 8092 is not satisfactory.} \label{Plot4Lav3-80lilo}
\end{figure}

      \begin{figure}
  \includegraphics[width=5.8in]
  {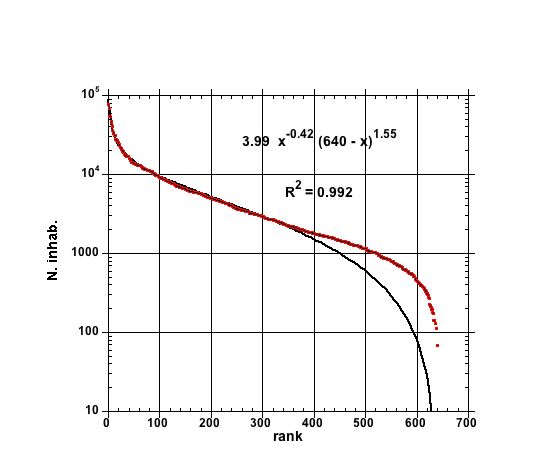} 
\caption{Blow up of the low rank of 8092 cities, Lavalette
3-parameters function. One can see some twisting effect of the
function due to the low rank outliers.} \label{Plot1Lav3lowrlilo}
\end{figure}

        \begin{figure}
  \includegraphics[width=5.8in]
  {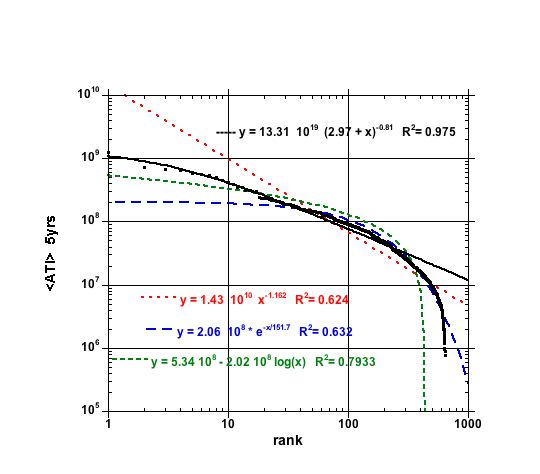} 
\caption{   Log-log display of the  the average ATI over 5 years for
the 639  (637 + 2; see text) Saints generating Italian
hagiotoponyms,  -  cities ranked in decreasing order of the ATI;
four simple law fits are shown with their corresponding regression
coefficient: power, exponential, log-, and BZM, Eq.(\ref{ZMlikeCr}).
\newline
Since these  trial fits with simple empirical laws  in  the "all
Italian Saints"  case,  are not  convincing  enough to represent the
data, they suggest to pursue further a data decomposition between
males and females.} \label{fig8b:Plot6ATI5lolo639fits4}
\end{figure}

        \begin{figure}
  \includegraphics[width=5.8in]
  {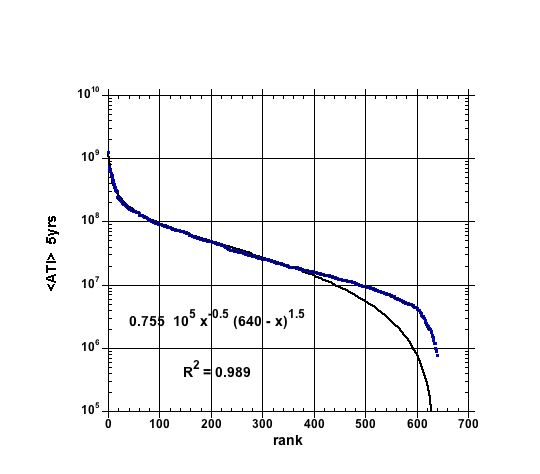} 
\caption{  Semi-log display of the  the average ATI over 5 years for
the 639  (637 + 2; see text)  Saints generating Italian
hagiotoponyms,   cities ranked in decreasing order of their ATI;   a
fit by a Lavalette function shows a convincing fit for $r\le350$.}
\label{figA2:Plot6ATI5lilo639fitsLav1}
\end{figure}

  \clearpage

        \begin{figure}
  \includegraphics[width=5.8in]   {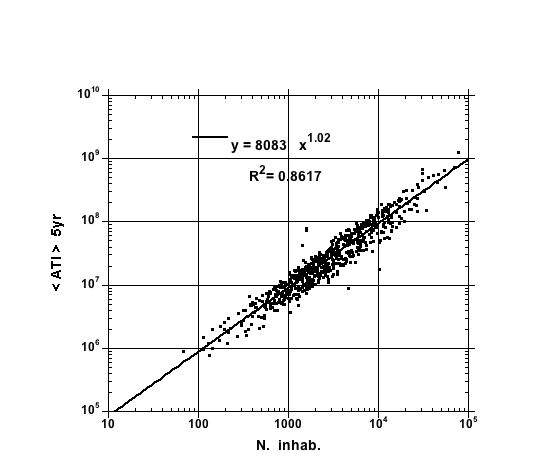}
\caption{   Log-log display of the   scatter plot for the averaged
over 5 years ATI and the number of inhabitants for   the 639 (637 +
2; see text)  Saints generating Italian hagiotoponyms, with the best
power law fit.} \label{fig11a:Plot1scatterNinhATIlolopw}
\end{figure}

        \begin{figure}
  \includegraphics[width=5.8in]
  {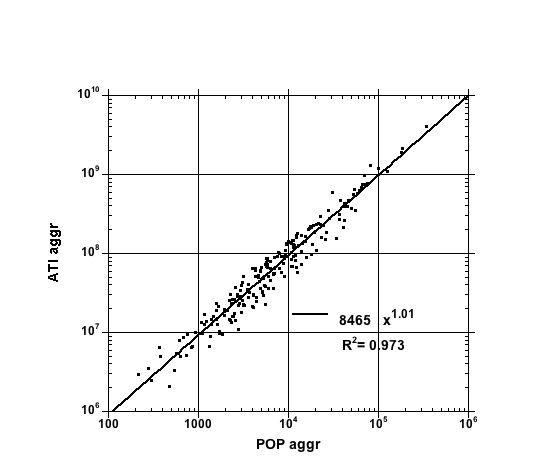}
\caption{   Log-log display of the  scatter plot of the  cumulated
averaged over 5 years  ATI and the cumulated Number of inhabitants
for the 206 different Saints,    with the best power law fit.}
\label{fig11b:Plot2scattNinhagATIaglolopw}
\end{figure}

        \begin{figure}
  \includegraphics[width=5.8in]
  {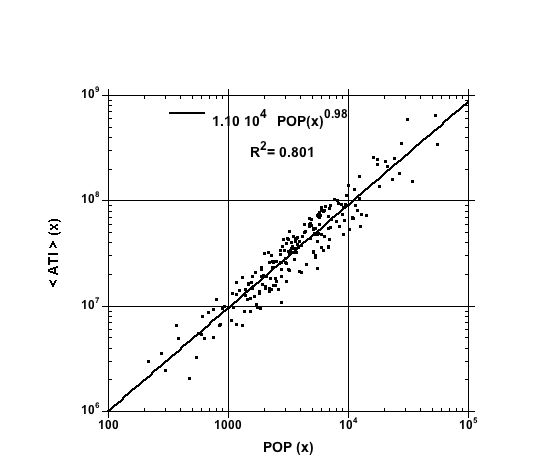}
\caption{Log-log display of  the  scatter plot of the averaged over
5 years  ATI $\overline{ATI}(x)$ and the Number of inhabitants
$\overline{POP}(x)$ for the 206 different Saints, reduced at the
frequency (popularity) of the Saint, with the best power law fit.}
\label{fig11c:Plot3scattNinhxATIxlolopw}
\end{figure}

        \begin{figure}
  \includegraphics[width=5.8in]   {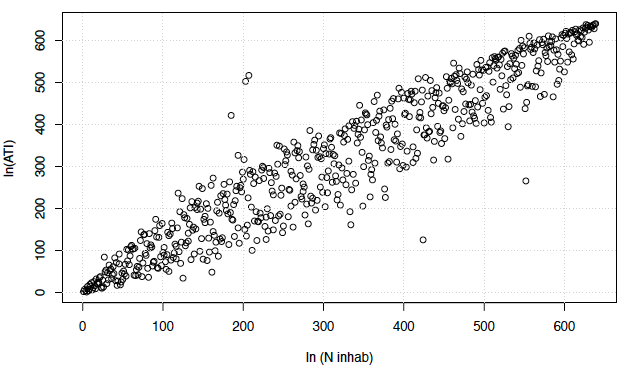}
\caption{    Log-log display of the   scatter plot of ranks for the
averaged over 5 years ATI and the Number of inhabitants for   the
639 Saints generating Italian hagiotoponyms.}
\label{fig11d:Plot1scatterNinhATIlolopw}
\end{figure}

        \begin{figure}
  \includegraphics[width=5.8in]
 {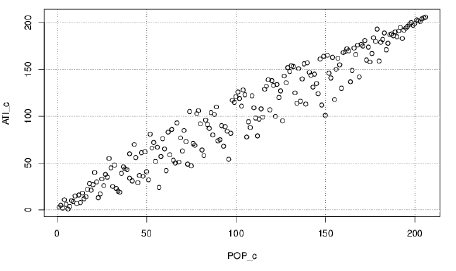}  
\caption{   Display of the  scatter plot of ranks for  $ATI(x)$,
i.e. the  cumulated averaged over 5 years  ATI, and the cumulated
number of inhabitants ($POP(x)$) for the 206 different Saints.}
\label{fig11e:Plot2scattNinhagATIaglolopw}
\end{figure}

        \begin{figure}
  \includegraphics[width=5.8in]
  {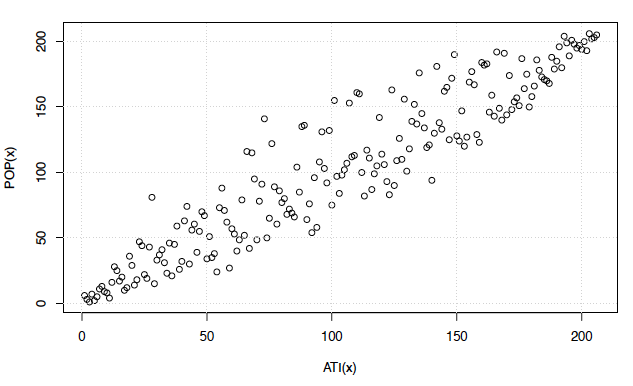}
\caption{   Display of  the  scatter plot of  ranks for  $ATI(x)$,
i.e. the averaged over 5 years  ATI and the number of inhabitants
($POP(x)$)  for   the 206 different Saints,    reduced by  the
frequency (popularity) of the Saint, i.e. $\overline{POP}(x)$ and
$\overline{ATI}(x)$.} \label{fig11f:Plot3scattNinhxATIxlolopw}
\end{figure}

 \end{document}